\documentclass[aps,prc,twocolumn,superscriptaddress,showpacs,showkeywords]{revtex4-1}

\usepackage{graphicx}
\usepackage{dcolumn}
\usepackage{subfigure}
\usepackage[T1]{fontenc}
\usepackage{mathptmx}

\begin{document}


\title{Transfer reactions in inverse kinematics, an experimental approach for fission investigations}


\author{C.~Rodr\'iguez-Tajes}
\email[]{carme.rodriguez@ganil.fr}
\affiliation{GANIL, CEA/DSM-CNRS/IN2P3, F-14076 Caen, France}\affiliation{Universidade de Santiago de Compostela, E-15706 Santiago de Compostela, Spain}

\author{F.~Farget}\email[]{fanny.farget@ganil.fr}\affiliation{GANIL, CEA/DSM-CNRS/IN2P3, F-14076 Caen, France}
\author{X. Derkx}\thanks{Present address: Institut f\"ur Kernchemie, Johannes Gutenberg-Universit\"at Mainz, D-55128 Mainz, Germany. Helmholtz Institut Mainz, D-5099 Mainz, Germany.}
\affiliation{GANIL, CEA/DSM-CNRS/IN2P3, F-14076 Caen, France}\affiliation{LPC, ENSICAEN, Universit\'e de Caen Basse Normandie, CNRS/IN2P3-ENSI, F-14050, France }
\author{M. Caama\~no}\affiliation{Universidade de Santiago de Compostela, E-15706 Santiago de Compostela, Spain}
\author{O. Delaune}\thanks{Present address: CEA, DAM, DIF, F-91297 Arpajon, France.}\affiliation{GANIL, CEA/DSM-CNRS/IN2P3, F-14076 Caen, France}
\author{K.-H. Schmidt}\affiliation{GANIL, CEA/DSM-CNRS/IN2P3, F-14076 Caen, France}
\author{E. Cl\'ement}\affiliation{GANIL, CEA/DSM-CNRS/IN2P3, F-14076 Caen, France}
\author{A. Dijon}\thanks{Present address: CEA, DAM, DIF, F-91297 Arpajon, France.}\affiliation{GANIL, CEA/DSM-CNRS/IN2P3, F-14076 Caen, France}
\author{A. Heinz}\affiliation{Fundamental Fysik, Chalmers Tekniska H\"ogskola, SE-41296 G\"oteborg, Sweden}
\author{T. Roger}\affiliation{GANIL, CEA/DSM-CNRS/IN2P3, F-14076 Caen, France}
\author{L. Audouin}\affiliation{Institut de Physique Nucl\'eaire, Universit\'e Paris-Sud 11, CNRS/IN2P3, F-91406, Orsay, France}
\author{J. Benlliure}\affiliation{Universidade de Santiago de Compostela, E-15706 Santiago de Compostela, Spain}
\author{E. Casarejos}\affiliation{Universidade de Vigo, E-36310 Vigo, Spain}
\author{D. Cortina}\affiliation{Universidade de Santiago de Compostela, E-15706 Santiago de Compostela, Spain}
\author{D.~Dor\'e}\affiliation{CEA, Irfu, Centre de Saclay, F-91191 Gif-sur-Yvette, France}
\author{B. Fern\'andez-Dom\'inguez}\affiliation{Universidade de Santiago de Compostela, E-15706 Santiago de Compostela, Spain}
\author{B. Jacquot}\affiliation{GANIL, CEA/DSM-CNRS/IN2P3, F-14076 Caen, France}
\author{B. Jurado}\affiliation{CENBG, UMR 5797 CNRS/IN2P3, Universit\'e Bordeaux, F-33175 Gradignan, France}
\author{A. Navin}\affiliation{GANIL, CEA/DSM-CNRS/IN2P3, F-14076 Caen, France}
\author{C. Paradela}\affiliation{Universidade de Santiago de Compostela, E-15706 Santiago de Compostela, Spain}
\author{D. Ramos}\affiliation{Universidade de Santiago de Compostela, E-15706 Santiago de Compostela, Spain}
\author{P. Romain}\affiliation{CEA, DAM, DIF, F-91297 Arpajon, France}
\author{M.D. Salsac}\affiliation{CEA, Irfu, Centre de Saclay, F-91191 Gif-sur-Yvette, France}
\author{C. Schmitt}\affiliation{GANIL, CEA/DSM-CNRS/IN2P3, F-14076 Caen, France}

\date{\today}

\begin{abstract}
Inelastic and multi-nucleon transfer reactions between a $^{238}$U beam, accelerated at 6.14 MeV/u, and a $^{12}$C target were used for the production of neutron-rich, fissioning systems from U to Cm. A Si telescope, devoted to the detection of the target-like nuclei, provided a characterization of the fissioning systems in atomic and mass numbers, as well as in excitation energy. Cross-sections, angular and excitation-energy distributions were measured for the inelastic and transfer channels. Possible excitations of the target-like nuclei were experimentally investigated for the first time, by means of $\gamma$-ray measurements. The decays from the first excited states of $^{12}$C, $^{11}$B and $^{10}$Be were observed with probabilities of 0.12 -- 0.14, while no evidence for the population of higher-lying states was found. Moreover, the fission probabilities of $^{238}$U, $^{239}$Np and $^{240,241,242}$Pu and $^{244}$Cm were determined as a function of the excitation energy.
\end{abstract}

%
%
%
%


\maketitle

\section{Introduction}

Multi-nucleon transfer reactions have been widely used for the investigation of the fission process, generally in direct-kinematics experiments with projectiles lighter than He and actinide targets. Good examples are the measurements of transfer-induced fission probabilities, which were used in the past as an experimental observable for the study of actinide fission barriers \cite{Gav76}.\\
In the framework of the surrogate-reaction technique, which is discussed in Sec.~\ref{sec:surrogate}, transfer-induced fission-probabilities allow for the estimation of neutron-induced fission cross-sections when direct neutron-irradiation measurements are not feasible \cite{Cra70,Pet04,Kes10,Bas09}. These measurements are important for nuclear energy applications, such as the development of new generation nuclear reactors or the recycling of radioactive waste.\\
In addition, the use of alternative reactions for fission investigations allows to extend the number of fissioning systems accessible to the experimental research and the investigation of fundamental properties of the fission process \cite{Sch00,Per07,And10,Bou13}.\\
Valuable results on fission probabilities were obtained by transfer-induced fission involving heavier projectiles, which are consistent with measurements using lighter projectiles and enlarged the accessible excitation-energy range \cite{Che81}. Furthermore, experiments of this type have permitted the investigation of some aspects of fission dynamics. In particular, high excitation energies and high angular momenta \cite{Vid77,Ger01} of the fissioning systems can be explored by increasing the charge and the mass transferred. Significant survival probabilities against fission have been observed, which give hints on the description of the fission times and the deexcitation of the compound nucleus \cite{Sax02}.\\
The interplay between fission and survival probabilities of the produced nuclear species brings particular interest to transfer-induced fission experiments where the fission of heavy or superheavy elements may be investigated. Actually, multi-nucleon transfer is expected to allow the production of more neutron-rich nuclei with longer half lives \cite{Zag13}.\\
Our experimental approach uses inelastic and multi-nucleon transfer reactions between a $^{238}$U beam and a $^{12}$C target for the production of the fissioning systems of interest, pushing the transfer-induced fission method towards heavier transfer reactions. In this way, a single experiment gives access to a higher variety of neutron-rich actinides and allows to explore different excitation energy regimes, depending on the transferred nucleons. In addition, the use of a heavy beam and a light target define an inverse-kinematics scenario, in which the fission fragments are emitted in forward direction with relatively high kinetic energies. Using a magnetic spectrometer, the accurate isotopic identification of the heavy and light fission fragments is possible. Results regarding isotopic fission-fragment yields from an earlier experimental campaign, can be found in Refs.~\cite{Caa13,Del_PhD}.\\
The present work aims to investigate the potential of $^{238}$U+$^{12}$C transfer reactions for fission investigations, by providing a detailed characterization of the different transfer channels and a discussion of the experimental data in the framework of the surrogate-reaction technique. The challenges associated with the detection of the relatively heavy target-like nuclei at high intensities and the use of inverse kinematics will be discussed. The latter, while improving the quality of the identification of the fission-fragments \cite{Sch00}, results in a degradation of the excitation-energy resolution. However, it also ensures cleaner experimental conditions than direct kinematics, where reactions on the backing of the actinide targets usually complicate the analysis of the experimental data and, in particular, the measurement of fission probabilities \cite{Pet04}.\\
The experimental set-up and data analysis are discussed in Secs.~\ref{sec:set-up} and \ref{sec:analysis}. Section \ref{sec:characterization} is dedicated to the characterization of $^{238}$U+$^{12}$C channels, providing cross-sections, angular and total excitation-energy distributions. The excitation of the target-like transfer partners is discussed in Sec.~\ref{sec:gamma}, in the light of $\gamma$-ray measurements. These results are especially important for surrogate-reaction fission experiments, as the excitation of target-like nuclei influences the excitation energy of the fissioning system. Fission probabilities are given in Sec.~\ref{sec:Pf} for $^{238}$U, $^{239}$Np, $^{240,241,242}$Pu and $^{244}$Cm, as a function of the total excitation energy. The latter is a short-lived ($T_{1/2}=18$ y) minor actinide, for which transmutation in accelerator-driven systems is nowadays a subject of recognized interest \cite{Art99}. Experimental data regarding fast-neutron induced fission of $^{244}$Cm are limited to Ref.~\cite{Fur97}. For the first time, this work brings access to excitation energies above 15 MeV for this particular nucleus. Finally, a summary and conclusions are presented in Sec.~\ref{sec:summary}. 

\subsection{The surrogate-reaction technique}
\label{sec:surrogate}

Following the Bohr hypothesis \cite{Boh36}, the surrogate-reaction method considers the formation and decay of the compound nucleus independent of each other. Therefore, in order to obtain information on its decay process, the compound nucleus of interest can be produced via a surrogate reaction, which is experimentally accessible. In fission investigations, inelastic and transfer reactions between nuclei in the actinide region and light nuclei are typically used.\\
The majority of surrogate applications invoke approximations, such as the Weisskopf-Ewing limit, where the decay branching ratios for the compound nucleus, formed with certain excitation energy, angular momentum and parity, are only a function of the excitation energy; or assume that both the desired and the surrogate reactions populate similar angular-momentum distributions in the compound nucleus \cite{Esc06,Esc12}. Under these conditions, the neutron-induced fission cross-section, $\sigma_{nf}$, can be estimated as the product of the fission probability measured in the surrogate reaction, $P_f$, and the compound-nucleus formation cross-section, $\sigma_n^{CN}$, which is calculated via an optical potential, i.e.

\begin{equation}
\sigma_{nf}(E_n) = \sigma_n^{CN}(E_n)P_f(E_x). 
\label{eq:surrogate}
\end{equation}

In this expression, $E_n$ represents the kinetic energy of the incident neutron and $E_x$ is the excitation energy of the compound nucleus. Both quantities are related through the neutron separation energy and the mass number of the compound nucleus, which are denoted as $S_n^{CN}$ and $A^{CN}$ in the equation below:
 
\begin{equation}
E_x  = E_n\frac{A^{CN}-1}{A^{CN}}+S_n^{CN}.
\label{eq:En}
\end{equation}
 
Although a good agreement has been observed between surrogate results and measured neutron-induced fission cross-sections \cite{Pet04,Kes10,Bas09}, their theoretical description has been the subject of intense investigation \cite{Rom12}, as the method is, in principle, restricted to specific conditions. The Weisskopf-Ewing approximation is only justified for high excitation energies, where the decay of the compound nucleus is dominated by statistical level densities, which have no dependence on the spin and parity \cite{Esc06}. In addition, different angular-momentum distributions of the produced compound nucleus are usually expected from transfer and neutron-capture reactions \cite{Esc12,Esc06,Bou12}. This conundrum becomes even more unclear due to the limited amount of available information on the angular-momentum of compound nuclei populated in transfer reactions, from both theoretical and experimental sides. Selecting the recoil angles of the surrogate reactions has been a method to show the influence of the angular momentum induced in the reaction on the fission-decay channel \cite{Lyl07}.\\  
It is also relevant to remark that the surrogate-reaction technique has been usually investigated in direct kinematics, using $^{1,2,3}$H, or $^{3,4}$He beams and actinide targets. The information concerning heavier transfer reactions, for which higher angular momenta are expected, is limited to previous measurements of $^{232}$Th($^{12}$C,$^8$Be)$^{236}$U and $^{236}$U($^{12}$C,$^{8}$Be)$^{240}$Pu \cite{Che81}, where a good agreement with direct neutron-induced fission cross-sections was obtained. A more recent $^{12}$C($^{238}$U,$^{240}$Pu)$^{10}$Be experiment at GANIL, in inverse kinematics \cite{Der10}, suffers from insufficient resolution in both the isotopic identification of the reaction channel and the determination of the excitation energy.\\
An additional issue in the context of inelastic- or transfer-induced, surrogate fission experiments lies in the determination of the actual excitation energy of the fissioning system. In standard surrogate-reaction measurements, the available excitation energy is usually attributed to the heavy transfer partner, i.e. the fissioning system. This is a fair consideration for reactions involving light nuclei where the first excited states are unbound or situated at high excitation energies, such as ($\alpha$,$\alpha^{\prime}$) \cite{Bur06}, or where the break-up of light transfer partner can be disregarded by a geometrical adjustment of the experimental set-up \cite{Gav76,Pet04}. However, in the perspective of a generalization of the method to heavier transfer reactions, its comprehensive application is more intricate. Besides the direct impact on the determination of the excitation energy at which fission occurs, the excited states of the light transfer partners eventually populated may decay by nucleon emission, leading to occasional misidentification of the transfer channels. In the present work, the light transfer partners present few bound states that decay through $\gamma$-ray emission. Their observation by means of $\gamma$-ray spectroscopy allows to investigate this challenging issue of the surrogate-reaction technique.

\section{Experimental set-up}
\label{sec:set-up}

\begin{figure}
\includegraphics[width=\linewidth]{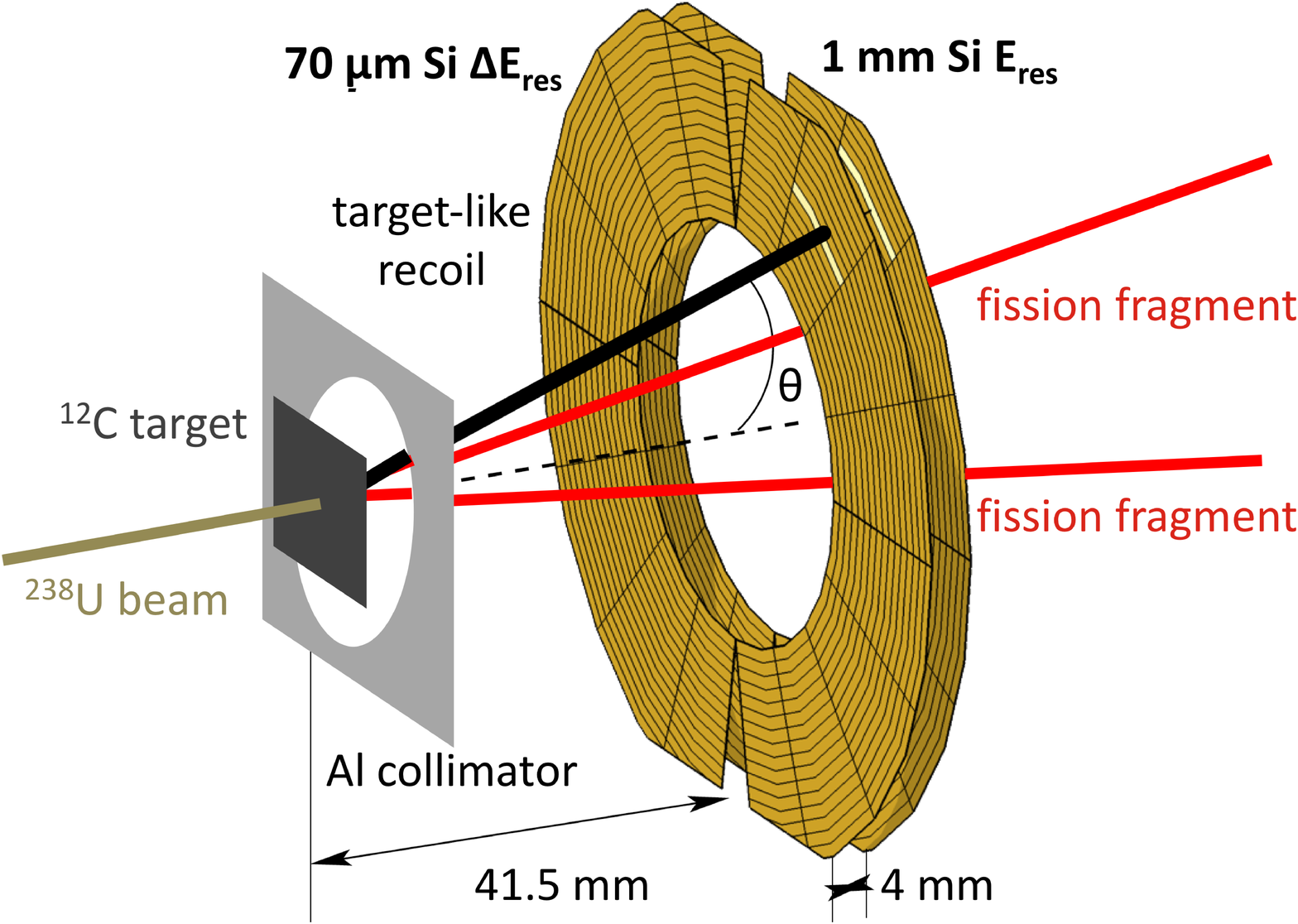}
\caption{\label{fig:spider}(Color online) Schematic layout of the SPIDER telescope for the detection of target-like nuclei. The beam and fission fragments passed through the inner hole of the telescope.}
\end{figure}

A $^{238}$U$^{31^+}$ beam with an average intensity of 10$^9$ pps was accelerated in the CSS1 cyclotron of GANIL up to 6.14 MeV/u. It impinged on a 100 $\mu$g/cm$^2$-thick $^{12}$C target. As a result, inelastic scattering, fusion and transfer reactions between the beam and the target were observed, producing a wide variety of excited actinides with a certain probability of decaying by fission. The incident energy in the center of mass, E$_{c.m.}$ = 70 MeV, was approximately 10 \% above the Coulomb barrier. In this scenario, as it has been shown in Ref.~\cite{Bis97}, fusion reactions, leading to the formation of $^{250}$Cf, dominate by far the total reaction cross-section, to which multi-nucleon transfer channels contribute with approximately 10 \%. The relatively low beam energy suppresses the opening of additional reaction channels.\\
The large-acceptance VAMOS spectrometer \cite{Rej11} was used for the identification of the fission fragments, described in detail in Refs.~\cite{Del12,Far12,Caa13}. In the present work, it was used as a fission-event detector.\\
The detection of the target-like nuclei was performed in a Si telescope named SPIDER, which is shown in Fig.~\ref{fig:spider}. It was located 41.5 mm behind the target and covered polar angles between 30$^{\circ}$, corresponding to the grazing angle \cite{Wil80}, and 47$^{\circ}$. The central hole of the detector ensured the non-interception of beam-like nuclei and fission fragments. In the inverse-kinematics conditions of the experiment, the former deviated only few degrees from the beam direction, while the latter were confined in a cone of about 25$^{\circ}$.\\
In order to avoid that $^{238}$U nuclei of the beam halo impinged on SPIDER, a 0.5 mm-thick Al collimator, with a radius of 4 mm, was placed 3.5 mm behind the target.\\
The results from a previous experiment showed an increase of the current in the SPIDER detector, which reached several $\mu$A, as a consequence of the high counting rates, up to 40 kHz, of high-energy elastically scattered $^{12}$C target nuclei. Therefore, the depletion of SPIDER was not permanently complete and its response dropped as a function of time. In addition, this earlier experience showed that the high counting rates increased the temperature of the detector, deteriorating the energy resolution. These limitations were overcome in the present work by using new preamplifier concepts and a cooling system, based on the circulation of liquid silicone at -30$^{\circ}$C, which avoided the degradation of the energy resolution. Details are provided in Ref.~\cite{Der_PhD}. Moreover, a magnetic field of approximately 750 G was used in the target region in order to prevent the arrival of $\delta$ electrons, produced in the interaction of the highly-charged $^{238}$U beam with the target, to SPIDER. \\
As shown in Fig.~\ref{fig:spider}, SPIDER is composed of two double-sided Si detectors, which are 70 and 1042 $\mu$m thick. They were used to measure the energy loss, $\Delta E$, and the residual energy, E$_{res}$, of the target-like nuclei. The front and back sides of each Si detector are respectively segmented into 16 rings of 1.5 mm and 16 sectors, each covering an azimuthal, angular range of 11$^{\circ}$. Ring (sector) sides are coated with 0.1 $\mu$m-thick Al (0.3 $\mu$m-thick Au) dead layers. The angles of the target-like nuclei with respect to the beam direction were measured with an uncertainty below 1$^{\circ}$ thanks to the annular segmentation of the telescope.\\
In addition, three clovers of the EXOGAM array of Ge detectors \cite{Sim00} surrounded the target region. They were placed at backward angles, between 120$^{\circ}$ and 150$^{\circ}$, the distances between the clovers and the target being 140.5, 160 and 158.1 mm. They were used in this work to investigate $\gamma$-ray emissions from the target-like nuclei.\\
Finally, two main acquisition triggers were used. They corresponded to the detection of a target-like nucleus in SPIDER, in coincidence and anticoincidence with the detection of a fission fragment in VAMOS. A reduction factor of 600 was applied to the latter in order to reduce the amount of $^{12}$C($^{238}$U,$^{238}$U)$^{12}$C elastic events treated by the data-acquisition system.

\section{Data analysis}
\label{sec:analysis}

\begin{figure}
\includegraphics[width=\linewidth]{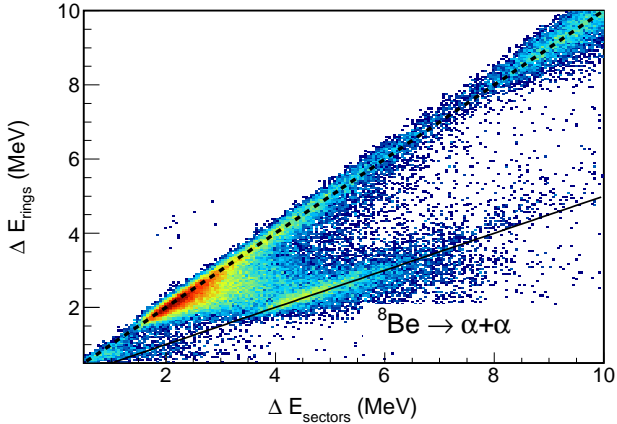}
\caption{\label{fig:mult2}(Color online) Correlation between ring and sector energy-loss measurements in the $\Delta E$ detector. Nuclei crossing a single ring of the $\Delta E$ detector are concentrated around the dashed line, for which $\Delta E_{rings}=\Delta E_{sectors}$. The solid line corresponds to $\Delta E_{rings}=0.5\cdot\Delta E_{sectors}$. Events in these region belong to the $^{12}$C($^{238}$U,$^{242}$Pu)$^{8}$Be channel.}
\end{figure}

\begin{figure}
\includegraphics[width=\linewidth]{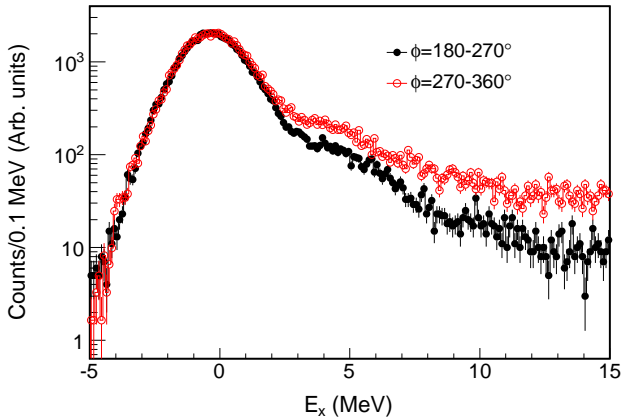}
\caption{\label{fig:Ex_mask}(Color online) Total excitation-energy distributions reconstructed for $^{12}$C($^{238}$U,$^{238}$U)$^{12}$C events. Empty (red) and solid (black) symbols correspond to different azimuthal angular ranges. The first distribution contains a higher proportion of wrongly reconstructed events due to scattering in the Al collimator.}
\end{figure}

The calibration of the SPIDER telescope was performed by means of $^{12}$C($^{238}$U,$^{238}$U)$^{12}$C elastic events. A beam energy of 6.11 MeV/u, which takes into account the slowing down in the first half of the target, was used for the kinematic calculations.\\
The rings and sectors of both $\Delta E$ and $E_{res}$ detectors were individually calibrated. Up to sixteen calibration points were obtained for each sector, using the coincidences with the associated rings.  For each ring, only one elastic point was available and we made also use of the pedestals.\\
These calibrations were found to be strongly influenced by geometric parameters, such as the position of the beam or the distance between the target and the telescope, which were adjusted using a detailed simulation of elastic events in our experimental set-up. This analysis showed that an accuracy of approximately 2 mm was achieved for the beam centering during the whole experiment. The beam size was of the same order, with FWHM of approximately 2.4 and 3.5 mm in the horizontal and vertical directions, respectively.\\ 
A double energy measurement was obtained in both $\Delta E$ and $E_{res}$ detectors, as the same information was provided by ring and sector signals. For each event, the ring and sector with the maximum deposited energy were selected. Geometrical considerations and the correlation between ring and sector energy measurements were used for the suppression of spurious events. Furthermore, the correlations between ring and sector energy measurements also determined the treatment of each event. Due to the thickness of the $E_{res}$ detector, some nuclei crossed several rings before being stopped. These events were identified by using ring-sector correlations, as the energy deposited in individual rings was smaller than that measured in the sectors. An add-back procedure between adjacent rings was applied in order to recover the total energy deposited from the ring signals.\\
The total kinetic energy of the target-like nuclei was finally obtained from SPIDER measurements as $E=\Delta E + E_{res} + \Delta E_{Al+Au}$. The term $\Delta E_{Al+Au}$ represents the energy loss in the Au and Al layers coating the detector sides and accounts for approximately 0.2--0.5 MeV. A precise estimation of $\Delta E_{Al+Au}$ was performed on an event-by-event basis, using the empirical formula given in ref.~\cite{Duf86}, in which the different parameters were adjusted by means of LISE++ simulations \cite{Tar01}.\\
In addition, the ring and sector energy measurements of the $\Delta E$ detector allowed to identify and to reconstruct multiplicity-two events arising from $^{12}$C($^{238}$U,$^{242}$Pu)$^{8}$Be reactions and the subsequent decay of $^{8}$Be into two strongly correlated $\alpha$ particles, with similar kinetic energies and a small angle between them. For those cases where the two $\alpha$ particles hit different rings and the same sector, a clear signature was found in the ring-sector energy correlation of the $\Delta E$ detector, which is illustrated in Fig.~\ref{fig:mult2}. The kinetic energy and the angle of $^{8}$Be nuclei were reconstructed from the kinetic energy, momentum and angle of the two $\alpha$ particles. They are denoted by $E_{\alpha_i}$, $P_{\alpha_i}$ and $\theta_{\alpha_i}$ in the equations below, where the angles are given with respect to the beam axis.

\begin{equation}
\begin{array}{c}
E_{^8Be} =E_{\alpha1}+E_{\alpha2}-\sqrt{(E_{\alpha1}+E_{\alpha2})^2-(\overrightarrow{P}_{\alpha1}+\overrightarrow{P}_{\alpha2})^2},\\
\\
cos(\theta_{^8Be}) =  \frac{ P_{\alpha_1}cos(\theta_{\alpha1}) + P_{\alpha_2}cos(\theta_{\alpha_2}) }{ \sqrt{(\overrightarrow{P}_{\alpha1}+\overrightarrow{P}_{\alpha2})^2}}. 
\end{array}
\label{eq:kin_mult2}
\end{equation}

The $\Delta E-E$ correlation was used to identify target-like nuclei according to their mass and atomic number, distinguishing the different transfer channels. The ring signals were used for this purpose because of their better energy resolution. The atomic and mass numbers of the complementary actinide partners were obtained assuming a binary reaction.\\
The measurement of the angle of the target-like nucleus with respect to the beam direction completed the reconstruction of the reaction kinematics. The trajectory of the beam was considered to be perpendicular to the target and well centered with respect to the SPIDER telescope. Then, the total excitation energy in the exit channel, $E_x$, was determined by applying energy and momentum conservation laws,

\begin{equation}
\begin{array}{c}
 E_{x}  =  Q_{gg} + E_{1} - E_{4} - \sqrt{M_3^2+p_3^{2}} + M_3,\\
\\
 p_3^2  =  p_1^2 + p_4^2 -2p_1p_4cos(\theta_4).
\end{array}
\end{equation} 

In these expressions, $Q_{gg}$ is the ground-state to ground-state Q-value of the reaction, $E_{i}$ represents the kinetic energy and the subindexes $i$=1,3 and 4 refer to the beam, the beam-like and the target-like nucleus, respectively. The angle of the target-like nucleus with respect to the beam direction is $\theta_4$. As shown in Fig.~\ref{fig:Ex}, an excitation-energy resolution of 2.7 MeV (FWHM) was achieved. The angular resolution, to which inverse-kinematics experiments are especially sensitive, was the main limiting factor.\\
The analysis of the experimental data revealed that the total excitation-energy distributions were affected by events where the target-like nuclei hit the inside of the Al collimator behind the target, which is represented in Fig.~\ref{fig:spider}. Nuclei scattered towards smaller polar angles in the collimator could reach the SPIDER telescope, biasing our reconstruction of the reaction kinematics and leading to an overestimation of the total excitation energy. For example, if $^{12}$C nuclei, elastically scattered from the target, were scattered 2$^{\circ}$ with respect to their original trajectories, total excitation energies of 6 MeV would be obtained. Figure \ref{fig:Ex_mask} provides evidence of this effect. The reconstructed total excitation-energy distributions are given for $^{12}$C nuclei detected at different azimuthal angular ranges, namely 180--270$^{\circ}$ and 270--360$^{\circ}$. The difference between the two distributions is caused by an inaccurate positioning of the Al collimator with respect to the beam axis, so target-like nuclei emitted at azimuthal angles of 270--360$^{\circ}$ had a higher probability of being scattered in the inside of the collimator. As a result, an important contribution from elastic events, wrongly reconstructed, was mixed with real inelastic events. The shadow of the collimator observed in the outer rings of SPIDER during the experiment gives further support to this hypothesis.\\
Consequently, only a reduced range of azimuthal angles was used for the data analysis in the present work, for which no shadow of the Al collimator was observed in the detector. Background contributions to the excitation energy spectra were reduced in this way by approximately a factor of four. An additional condition, which excluded the highest polar angles, above 40 deg, was applied for the determination of the total excitation-energy distributions and fission probabilities, discussed in Secs.~\ref{sec:Ex} and \ref{sec:Pf}, in order to supress residual background events.\\
The information provided by SPIDER was also used for the Doppler correction of the $\gamma$-ray energies performed with EXOGAM clovers. In this analysis, the target-like nuclei, moving with typical velocities of 5 cm/ns, were considered as the $\gamma$-ray emitters.\\
The efficiency of the $\gamma$-ray detection was evaluated in two steps. Firstly, the dependence on the $\gamma$-ray energy was determined from measurements performed during our experiment with a $^{152}$Eu calibration source and $^{35}$Cl(n,$\gamma$) data available from an earlier work \cite{Mol04,Mut13}, in which $\gamma$-ray energies of up to 9 MeV were reached. Absolute efficiency values were then obtained by applying a scaling factor, based on a $^{60}$Co calibration source located at the target position. A 20 Hz pulser was used in order to evaluate the dead-time contribution to the $^{60}$Co measurements, which was taken into account afterwards. Absolute efficiency values of 0.014 and 0.006 were obtained at energies of 1.3 and 4 MeV, respectively. An approximate error of 6 \% was estimated for the efficiency values obtained by this procedure.

\section{Characterization of $^{238}$U+$^{12}$C inelastic and transfer reactions}
\label{sec:characterization}

\subsection{Identification of the exit channels}
\label{sec:identification}

\begin{figure*}[t!]
\includegraphics[width=0.49\linewidth]{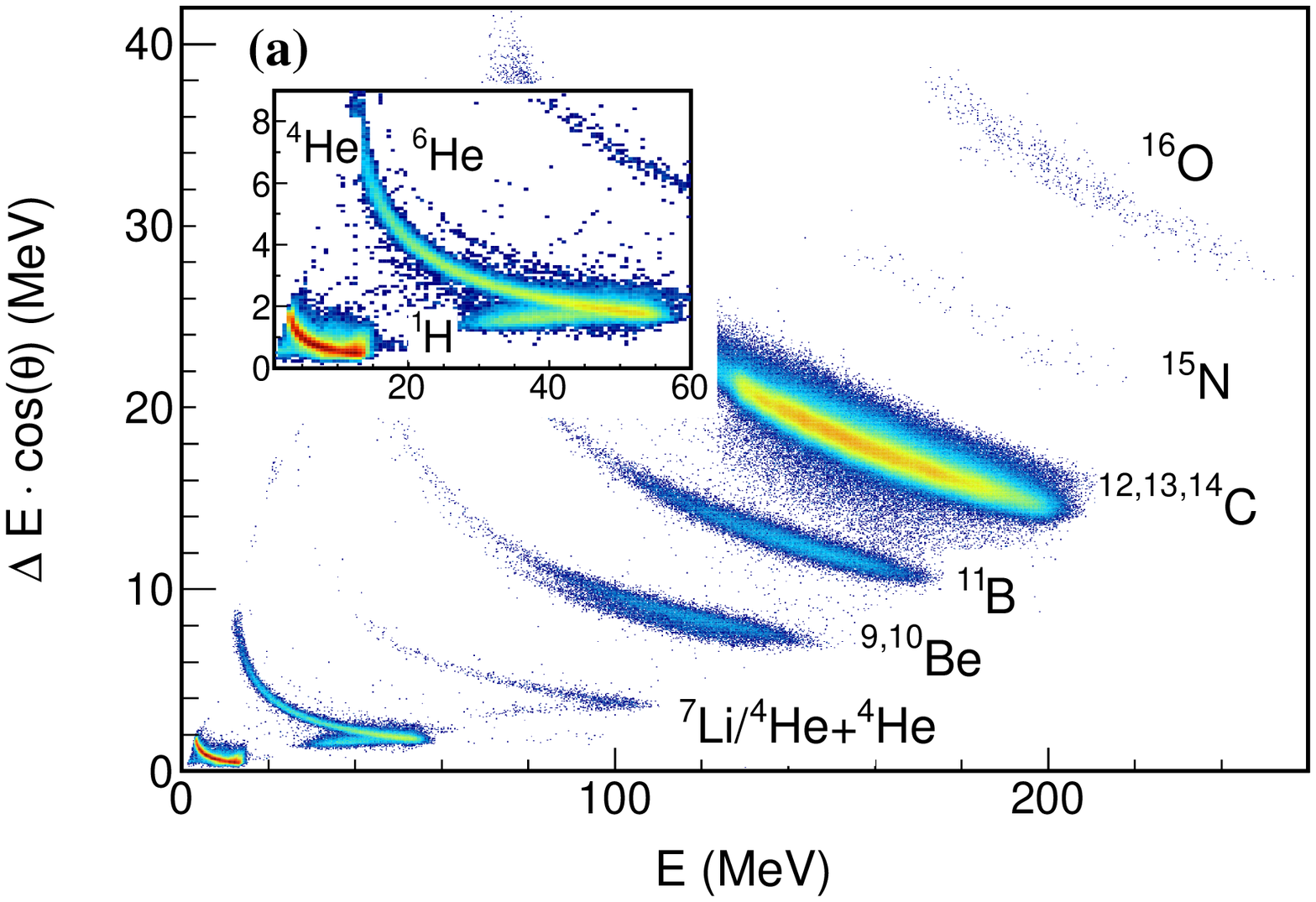}
\includegraphics[width=0.49\linewidth]{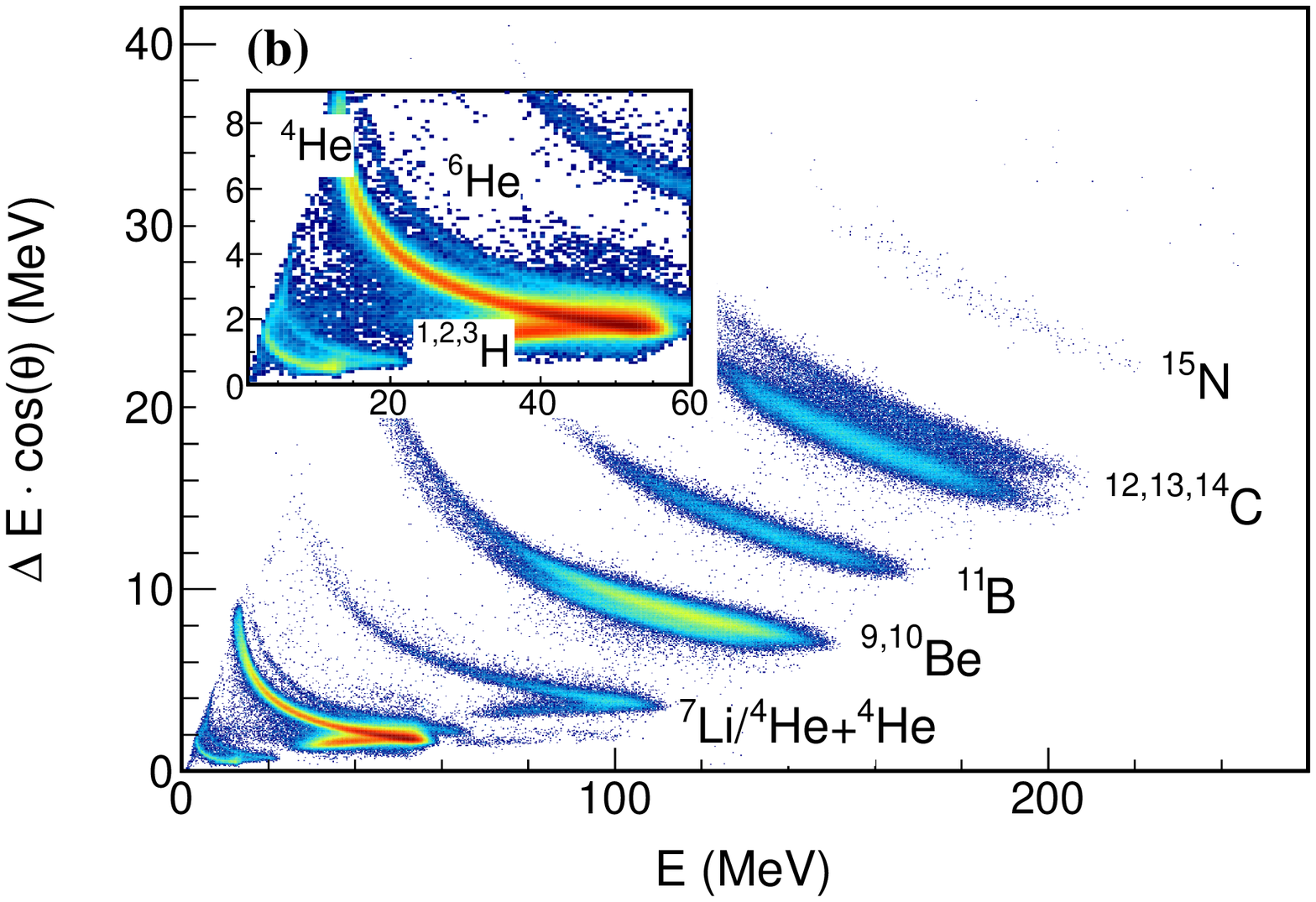}
\caption{\label{fig:id}(Color online) Identification of target-like nuclei in the SPIDER telescope. The energy loss in the $\Delta$E detector is plotted on the vertical axis. The factor $cos(\theta)$ accounts for the different effective thickness crossed by target-like nuclei emitted at different angles, $\theta$. The total kinetic energy is represented on the horizontal axis. Experimental data are shown in anticoincidence (a) and coincidence (b) with the detection of a fission fragment in VAMOS.}
\end{figure*}

\begin{figure*}[t!]
\includegraphics[width=0.49\linewidth]{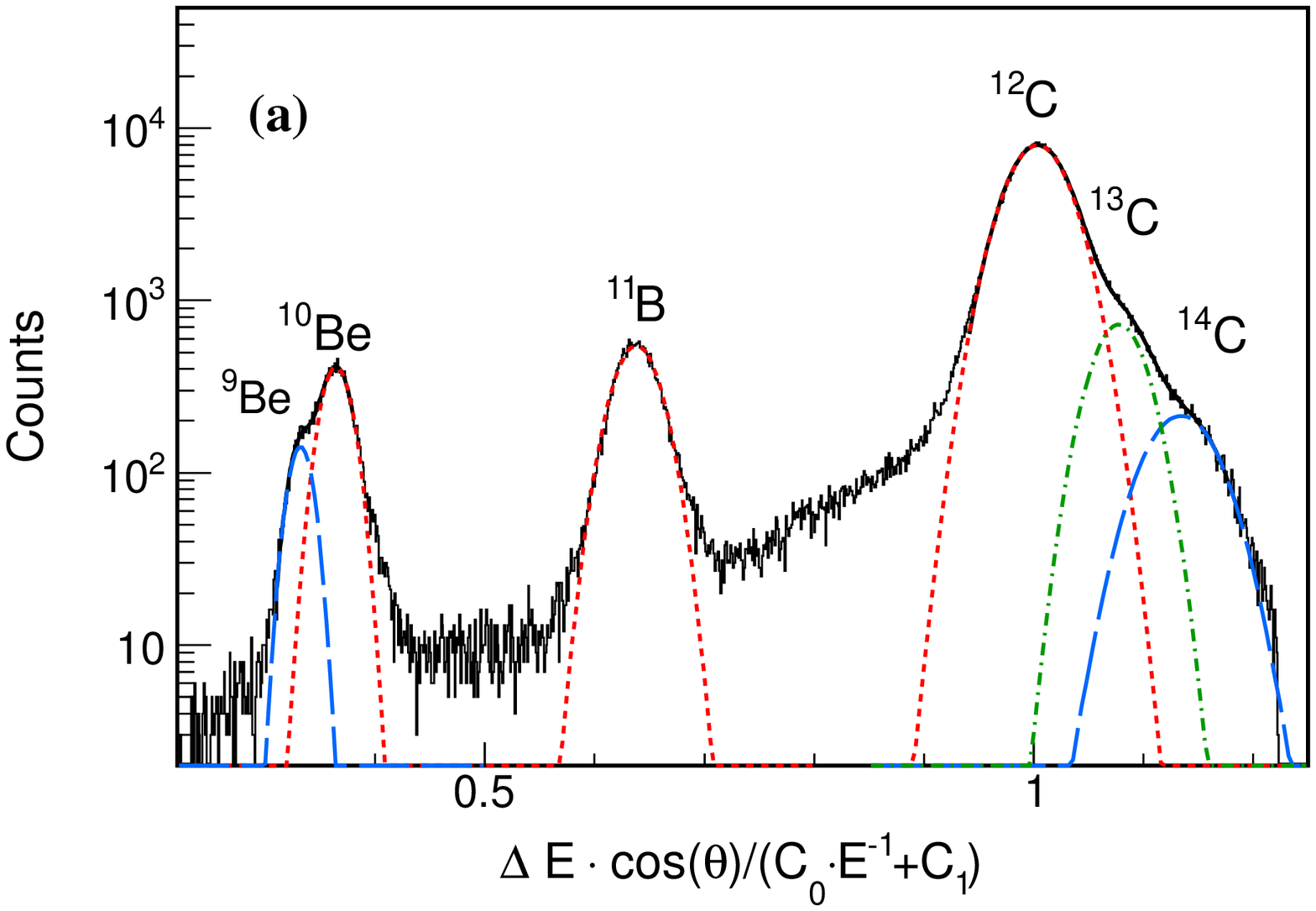}
\includegraphics[width=0.49\linewidth]{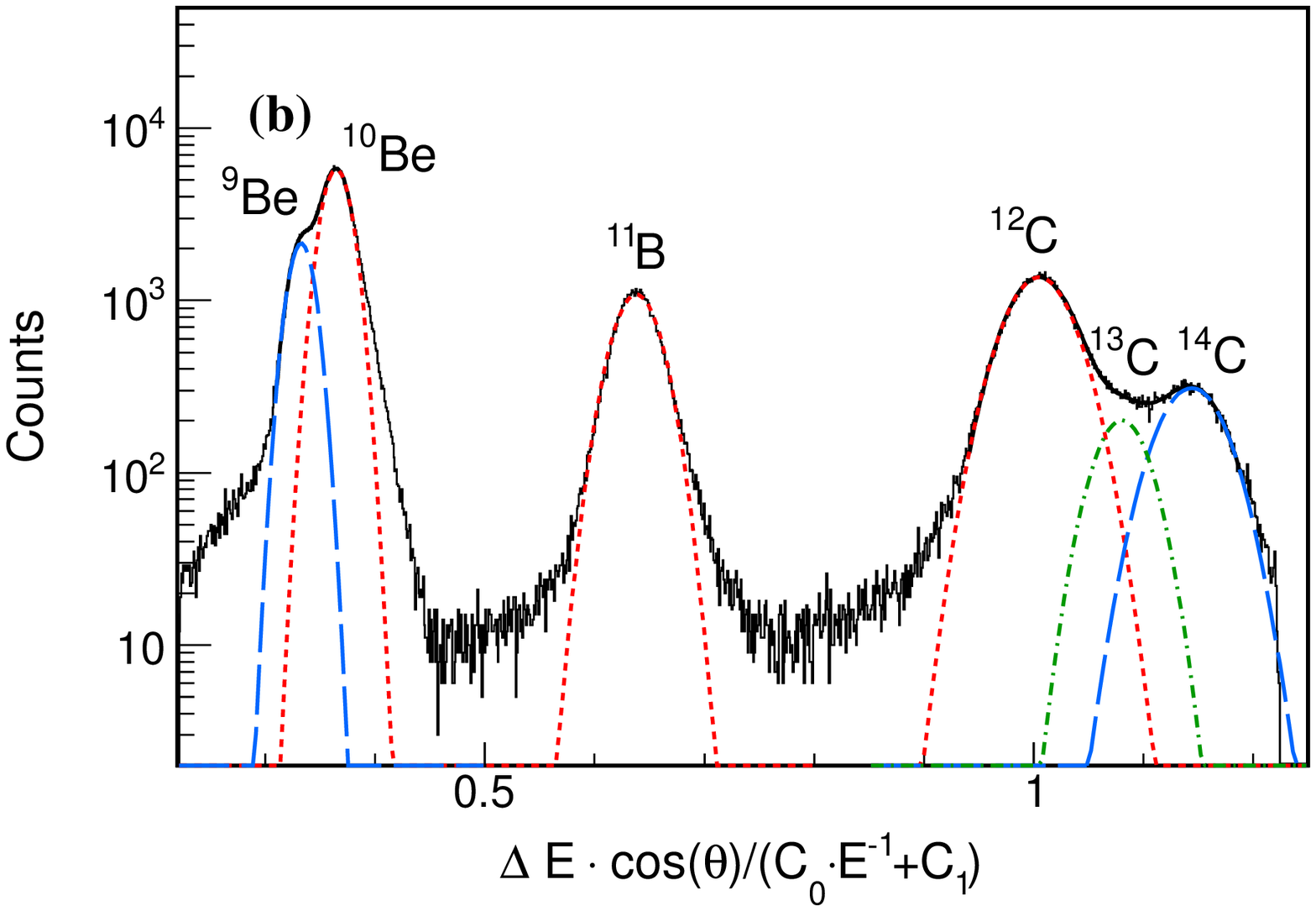}
\caption{\label{fig:IdE}Mass-separation of the target nuclei, in anticoincidence (a) and coincidence (b) with the detection of a fission fragment in VAMOS. The coefficients $C_{0,1}$ were determined by fitting the experimental data. The curves therein are the result of multiple-Gaussian fits, which were used to estimate the mixing between $^{9,10}$Be and $^{12,13,14}$C isotopes.}
\end{figure*}

\begin{table}
\caption{\label{tab:mixing_ID} Contributions from neighboring isotopes in the selection of target-like nuclei. The results are given in anticoincidence and coincidence with the detection of a fission fragment, $\overline{fission}$ and ${fission}$.}
\begin{ruledtabular}
\begin{tabular}{c c c}
        &  $\overline{fission}$ & ${fission}$ \\
$N_{^{12}C}/(N_{^{12}C}+N_{^{13}C})$    & 0.97 $\pm$ 0.02 & 0.97 $\pm$ 0.02 \\
$N_{^{14}C}/(N_{^{13}C}+N_{^{14}C})$    & 0.43 $\pm$ 0.12 & 0.79 $\pm$ 0.11 \\
$N_{^{9}Be}/(N_{^{9}Be}+N_{^{10}Be})$ & 0.71 $\pm$ 0.06 & 0.81 $\pm$ 0.09 \\
$N_{^{10}Be}/(N_{^{9}Be}+N_{^{10}Be})$ & 0.96 $\pm$ 0.02 & 0.95 $\pm$ 0.02
\label{tab:mixing}
\end{tabular}
\end{ruledtabular}
\end{table}

The identification of target-like nuclei provided by the SPIDER telescope is displayed in Fig.~\ref{fig:id}, where the energy loss in the $\Delta$E detector is shown as a function of the total kinetic energy, $E$. In this representation, the effective thickness crossed by target-like nuclei emitted at different angles, $\theta$, is accounted for by the factor $cos(\theta)$. Therefore, the quantity $\Delta E \cdot cos(\theta)$ depends only on the atomic number, the mass and the total kinetic energy. Results are given for two trigger conditions, which select the anticoincidence and the coincidence with the detection of a fission fragment in VAMOS, as mentioned above. The latter requires a minimum excitation energy of approximately 6 MeV, corresponding to the height of the fission barrier. Hence, different relative populations of the exit channels are observed in the two cases. The anticoincidences with the detection of a fission fragment are clearly dominated by elastic events, which are suppressed when a coincidence with a fission fragment is required.\\
The atomic and mass numbers assigned to the target-like nuclei were cross-checked with simulations of the energy loss in the SPIDER telescope. Graphical cuts on the identification of Fig.~\ref{fig:id} were used for the selection of the different reaction channels. Assuming a linear dependence between $\Delta E \cdot cos(\theta)$ and $E^{-1}$, the ratio $\frac{ \Delta E \cdot cos(\theta) }{C_0\cdot E^{-1}+C_1}$ is presented in Fig.~\ref{fig:IdE}, in order to illustrate the mass-separation achieved for the target-like nuclei. The coefficients $C_{0,1}$ were obtained from a fit of the experimental data. A mass resolution of approximately 8\% (FWHM) was obtained in this work.\\
Multiple-Gaussian fits of the spectra given in Fig.~\ref{fig:IdE} were used to estimate the mixing between $^{9,10}$Be and $^{12,13,14}$C isotopes. The contributions to the graphical cuts that we used to select the different target-like nuclei are given in Table \ref{tab:mixing_ID}.\\
The presence of $^{16}$O, $^{15}$N and $^{1}$H was attributed to $^{16}$O($^{238}$U,$^{238}$U)$^{16}$O, $^{16}$O($^{238}$U,$^{239}$Np)$^{15}$N and $^{1}$H($^{238}$U,$^{238}$U)$^{1}$H reactions, as a result of contaminants in the target. The fact that $^{16}$O and $^1$H were mostly detected in anticoincidence with a fission fragment, together with the angular distributions measured for these nuclei, suggests that they mainly emerged from elastic reactions. The kinematic $E$ vs. $\theta$ correlations observed for these nuclei firmly support this hypothesis. Due to the importance of $^1$H contaminants, the origin of the observed $^2$H and $^3$He remains unclear.\\
Nuclei from C down to He were assigned to $^{238}$U+$^{12}$C reactions, leading to the production of actinides between U and Cm. Besides elastic and inelastic scattering, up to nine transfer channels were populated. Our data indicate that nucleons essentially flow from the light to the heavy nucleus. The transfer in the opposite direction was restricted to one or two neutrons, leading to $^{13}$C and $^{14}$C target-like nuclei. The presence of $^{13}$C was confirmed from $\gamma$-ray spectra. Characteristic $\gamma$-rays at 3684 and 3853 keV, depopulating $^{13}$C excited states, are clearly visible in Fig.~\ref{fig:gamma} (a).\\
Furthermore, $^{11}$B, $^{10,9}$Be nuclei were observed. The transfer of a $^{4}$He nucleus lead to $^8$Be, which decayed into two highly-correlated $\alpha$ particles, its half-life being $T_{1/2}\sim8\cdot10^{-17}$ s. When they hit the same ring and sector of both $\Delta E$ and $E_{res}$ detectors, they populated the same $\Delta$E--E region as $^{7}$Li in the identification matrices of Fig.~\ref{fig:id}. Although transfer reactions leading to the production of $^8$Be are clearly favored by their ground-state to ground-state Q-value, possible $^7$Li contributions could not be disentangled. For reference, ground-state to ground-state Q-values of the observed channels are given in Table \ref{tab:xs_Ex}.\\
The observed $^6$He nuclei were attributed to $^{12}$C($^{238}$U,$^{244}$Cm)$^6$He. Transfer reactions leading to $^7$He, which would decay into n+$^6$He with $T_{1/2}\sim3\cdot10^{-21}$ s, were neglected, as they would be strongly disfavored by their ground-state to ground-state Q-value, $Q_{gg}$ = -35.95 MeV. \\
The observed $^4$He could originate from several channels. Multiplicity-two events, in which two $^4$He nuclei were detected by SPIDER, were unambiguously attributed to $^{12}$C($^{238}$U,$^{242}$Pu)$^8$Be reactions. An example is given in Fig.~\ref{fig:mult2}. However, those cases where a single $^4$He nucleus was detected represent a more complicated puzzle. They could be as well $^{12}$C($^{238}$U,$^{242}$Pu)$^8$Be events, in which one $\alpha$ particle was lost. Although, from a geometrical point of view, the detection of the two highly-correlated $\alpha$ particles was very probable, a limited detection efficiency for such a scenario was found during the experiment, which did not allow to exclude this hypothesis. Additional contributions could originate from $^{12}$C($^{238}$U,$^{246}$Cm)$^4$He reactions. The $^{12}$C($^{238}$U,$^{245}$Cm)$^5$He channel, where $^5$He would decay into $n$+$^4$He with $T_{1/2}\sim8\cdot10^{-22}$ s, is considered negligible, as it has $Q_{gg}$ = -24.93 MeV.

\subsection{Cross-sections and angular distributions}
\label{sec:cross-sections}

\begin{figure*}
\includegraphics[width=\linewidth]{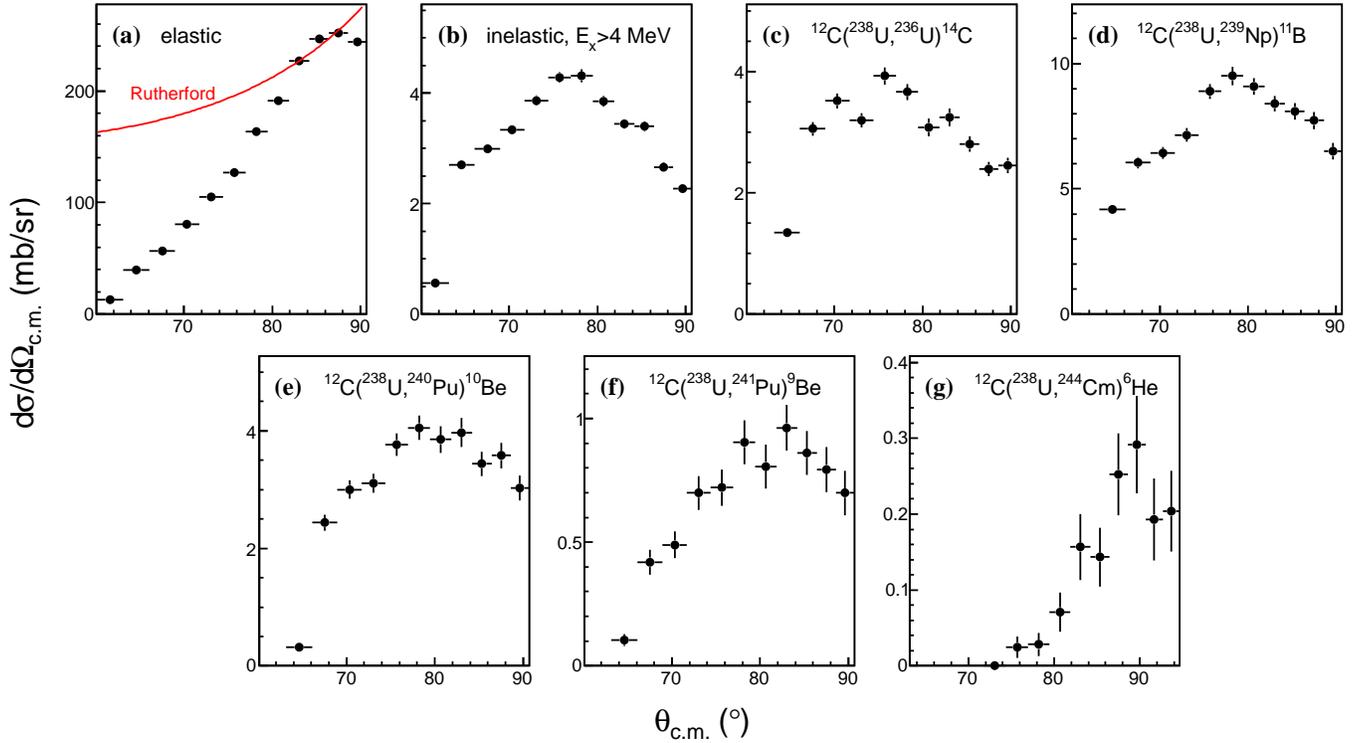}	
\caption{\label{fig:Th}(Color online) Differential cross-sections measured for $^{238}$U+$^{12}$C elastic, inelastic, and multi-nucleon transfer reactions. The angles of the target-like nuclei, in the center of mass reference frame, are plotted on the abscissa axis. The horizontal error bars reflect the ring size of the SPIDER $\Delta$E detector and the vertical error bars are statistical. In (a), the Rutherford cross-section is shown together with the elastic data, in order to illustrate the normalization applied to the experimental cross-sections.}
\end{figure*}

\begin{figure}
{\includegraphics[width=\linewidth]{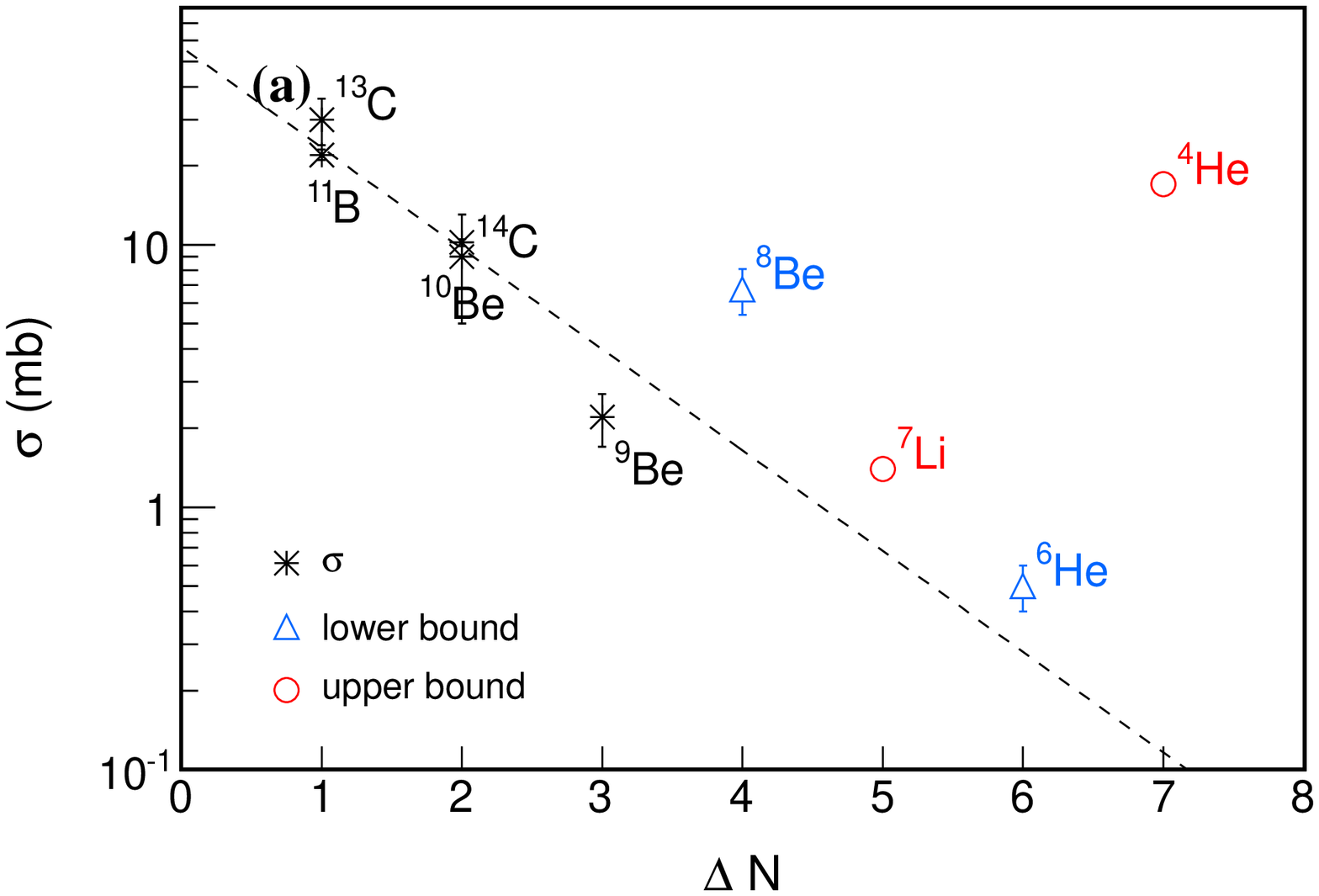}}
{\includegraphics[width=\linewidth]{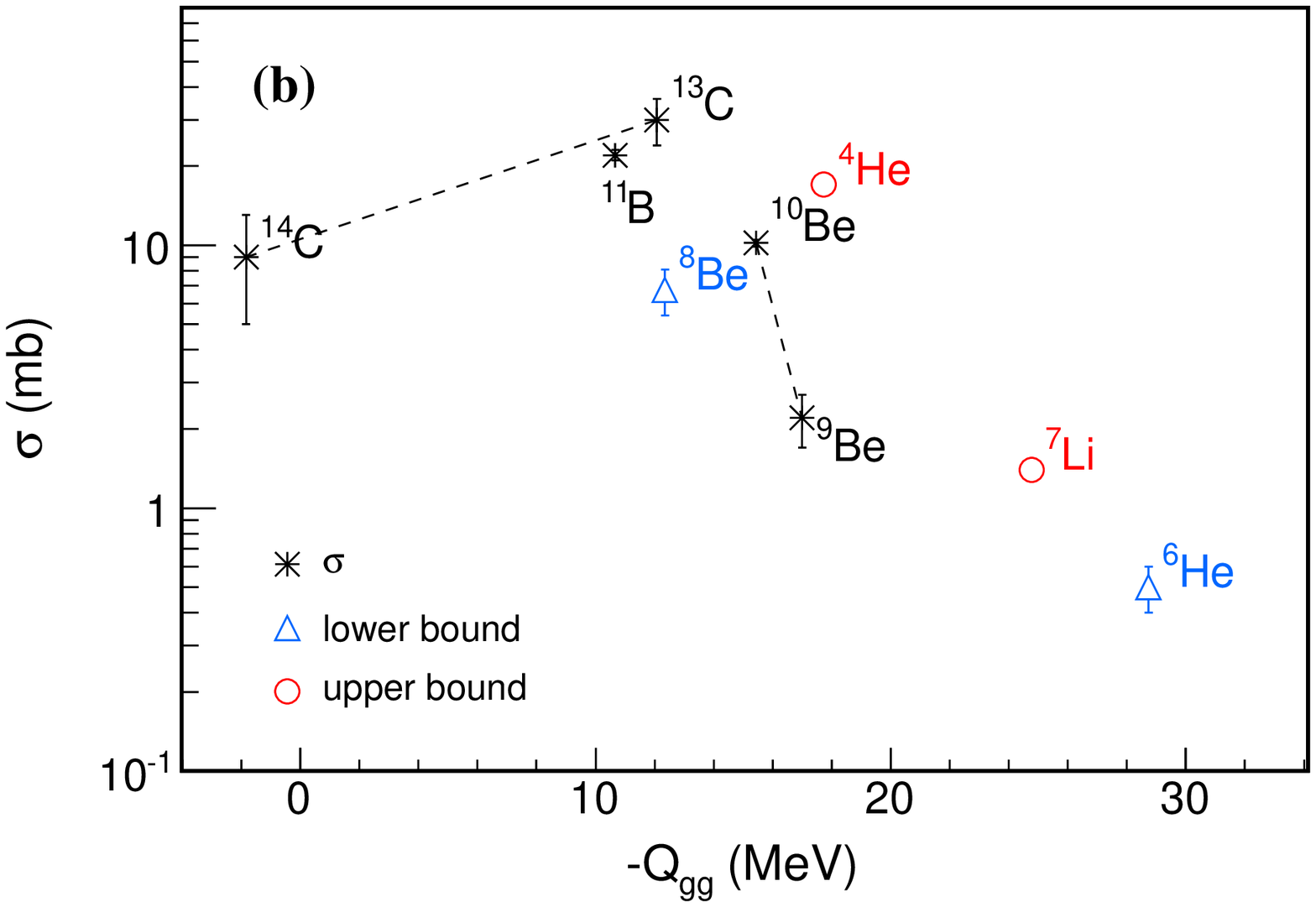}}
\caption{\label{fig:xs}(Color online) Variation of the transfer cross-sections with the number of transferred nucleons (a) and the ground-state to ground-state Q-value (b). In those cases where the actual cross-sections were not accessible, lower (upper) bounds are represented by empty triangles (circles). The dashed line in (a) corresponds to an exponential fit of the experimental data. The lines in (b) simply indicate the evolution of the cross-sections for C and Be isotopes.}
\end{figure}

\begin{table*}
\caption{\label{tab:xs_Ex}
Experimental $^{238}$U+$^{12}$C cross-sections, $\sigma^{exp}$, integrated over the angular ranges displayed in Fig.~\ref{fig:Th}. The most probable total excitation energies and the associated optimal Q-values are represented by $E_x^{exp}$ and $Q_{opt}^{exp}$, respectively.  For comparison, the optimal Q-values predicted by Eq.~\ref{eq:Qopt}, $Q_{opt}^{cal}$, are indicated, as well as the incident neutron energies leading to the reported $E_{x}^{exp}$ values in a neutron capture reaction, $E_n$. The ground-sate Q-values, $Q_{gg}$, are also indicated.}
\begin{ruledtabular}
\begin{tabular}{c d c c c c c}
Reaction &  \multicolumn{1}{c}{$Q_{gg}$ (MeV)} & $\sigma^{exp}$ (mb)& $E_{x}^{exp}$ (MeV)& $Q_{opt}^{exp}$ (MeV) & $Q_{opt}^{calc}$ (MeV) & \multicolumn{1}{c}{$E_{n}$ (MeV)}\\
Elastic                                                                     & 0.00   & $417\pm52$   &  --           & --            & 0.00 & --\\
Inelastic, $E_x>$4 MeV                                      & 0.00   & $16\pm2$     &  --           & --            & 0.00 & thermal  \\
$^{12}$C($^{238}$U,$^{237}$U)$^{13}$C    & -12.08 & $30\pm6$     &  --           & --            & 0.00 & -- \\
$^{12}$C($^{238}$U,$^{236}$U)$^{14}$C     & 1.85   & $9\pm4$      &  $5.3\pm0.8$  & $-3.4\pm0.1$  & 0.00 & thermal  \\
$^{12}$C($^{238}$U,$^{239}$Np)$^{11}$B   & -10.67 & $22\pm1$     &  $2.9\pm0.1$  & $-13.7\pm0.1$ & -11.00 & thermal  \\
$^{12}$C($^{238}$U,$^{240}$Pu)$^{10}$Be  & -15.43 & $10.2\pm0.3$ &  $9.9\pm0.1$  & $-25.3\pm0.1$ & -22.26 & 4 \\
$^{12}$C($^{238}$U,$^{241}$Pu)$^{9}$Be    & -17.00 & $2.2\pm0.5$  &  $10.0\pm0.2$ & $-26.9\pm0.1$ & -22.26 & 5   \\
$^{12}$C($^{238}$U,$^{242}$Pu)$^{8}$Be    & -12.35 & $>6.7\pm1.4$     &  $18.8\pm0.1$ & $-31.5\pm0.1$ & -22.26 & 12 \\
$^{12}$C($^{238}$U,$^{243}$Am)$^{7}$Li    & -24.77 & $<1.4\pm0.1$ &  --           & -- & -33.76 & -- \\
$^{12}$C($^{238}$U,$^{244}$Cm)$^{6}$He  & -28.74 & $>0.5\pm0.1$ &  $21.3\pm0.5$ & $-50.0\pm0.1$ & -45.52 & 14 \\
$^{12}$C($^{238}$U,$^{246}$Cm)$^{4}$He  & -17.73 & $<17\pm1$    &  --           & --& -45.52 & --  
\end{tabular}
\end{ruledtabular}
\end{table*}

Figure \ref{fig:Th} shows the differential cross-sections measured for $^{238}$U+$^{12}$C elastic, inelastic and multi-nucleon transfer reactions, as a function of the angle of the target-like nuclei, in the center of mass reference frame. The distinction between elastic and inelastic events was done by means of the reconstructed excitation energy, given in Fig.~\ref{fig:Ex}. A maximum excitation energy of 2 MeV was requested in the selection of the elastic channel. This limit corresponds to two times the width of the elastic peak, approximately. A correction factor of $\sim3\%$, which accounted for the elastic events beyond this value, was considered for the determination of the cross-section. In order to minimize possible contamination from the elastic channel, only inelastic events with excitation energies above 4 MeV were considered in this analysis.\\
As illustrated in Fig.~\ref{fig:Th}, the elastic data were normalized to the Rutherford formula at the highest scattering angles. On this basis, they were used to perform the beam normalization and obtain absolute cross-sections.\\
In general, the inelastic and transfer channels show similar bell shapes peaked between 75 and 80$^{\circ}$, in the center of mass. The maxima are above the calculated grazing angle, which has a value of 60$^{\circ}$ \cite{Wil80}. No clear evolution of the width of these distributions with the number of transferred nucleons was noted. Moreover, the distributions resulting from $^{12}$C($^{238}$U,$^{244}$Cm)$^6$He should be interpreted carefully. They do not include \emph{punch-through} events, associated with $^6$He nuclei that were not stopped in the SPIDER telescope, because they are mixed with $^4$He in the identification matrix of Fig.~\ref{fig:id} (inset). As these events correspond to the smallest $^6$He angles, their exclusion modifies the differential cross-section.\\
The integrated cross-sections corresponding to the angular ranges displayed in Fig.~\ref{fig:Th} are given in Table \ref{tab:xs_Ex}. They were corrected for the contamination between $^{9,10}$Be and $^{12,13,14}$C isotopes, which is represented in Fig.~\ref{fig:IdE} and Table \ref{tab:mixing}.\\
The lower bound given for the $^{12}$C($^{238}$U,$^{242}$Pu)$^8$Be reactions was obtained from multiplicity-two events, in which two $^4$He nuclei were simultaneously detected by SPIDER. In our calculations, all the events lying in the region of Fig.~\ref{fig:id} labeled as $^7$Li/$^4$He+${^4}$He were assigned to $^{12}$C($^{238}$U,$^{243}$Am)$^7$Li reactions. As possible contributions are expected from the $^{12}$C($^{238}$U,$^{242}$Pu)$^8$Be channel, incompletely reconstructed, our result must be regarded as an upper bound. In a similar way, the events associated with the detection of a single $^4$He nucleus were assigned to $^{12}$C($^{238}$U,$^{246}$Cm)$^{4}$He reactions, despite contaminants from other channels, such as $^{12}$C($^{238}$U,$^{242}$Pu)$^8$Be, are expected. Finally, the amount of $^6$He nuclei that were not stopped in SPIDER could not be estimated. Therefore, the reported $^{12}$C($^{238}$U,$^{244}$Cm)$^6$He cross-section should be considered as a lower limit.\\ 
The ensemble of the measured cross-sections are represented in Fig.~\ref{fig:xs} (a) as a function of the number of transferred nucleons. In general, they decrease exponentially with the number of transferred nucleons, following the systematics observed in earlier works \cite{Kar82}. The transfer of two neutrons and two protons, which leads to $^{8}$Be and $^{242}$Pu, is favored by at least a factor of four with respect to the reported tendency, reflecting the clustering of the transferred nucleons into an $\alpha$ particle. Similar results were found in Refs.~\cite{Bis95,Kar82} for different projectile-target systems.\\
The evolution of the cross-sections with the ground-state to ground-state Q-value, $Q_{gg}$, is displayed in Fig.~\ref{fig:xs} (b). An exponential decrease with $-Q_{gg}$ has been observed in earlier works \cite{Bis95,Kar82}, although large deviations have been found and the need of a Coulomb correction term has been pointed out \cite{Kar82}. Systematics for the different charges are limited in our case to $^{13,14}$C and $^{8,9,10}$Be isotopes. The first clearly confirms that the number of transferred nucleons is a more appropriate ordering parameter for the description of the cross-sections. 

\subsection{Total excitation-energy distributions}
\label{sec:Ex}

\begin{figure*}
\includegraphics[width=\linewidth]{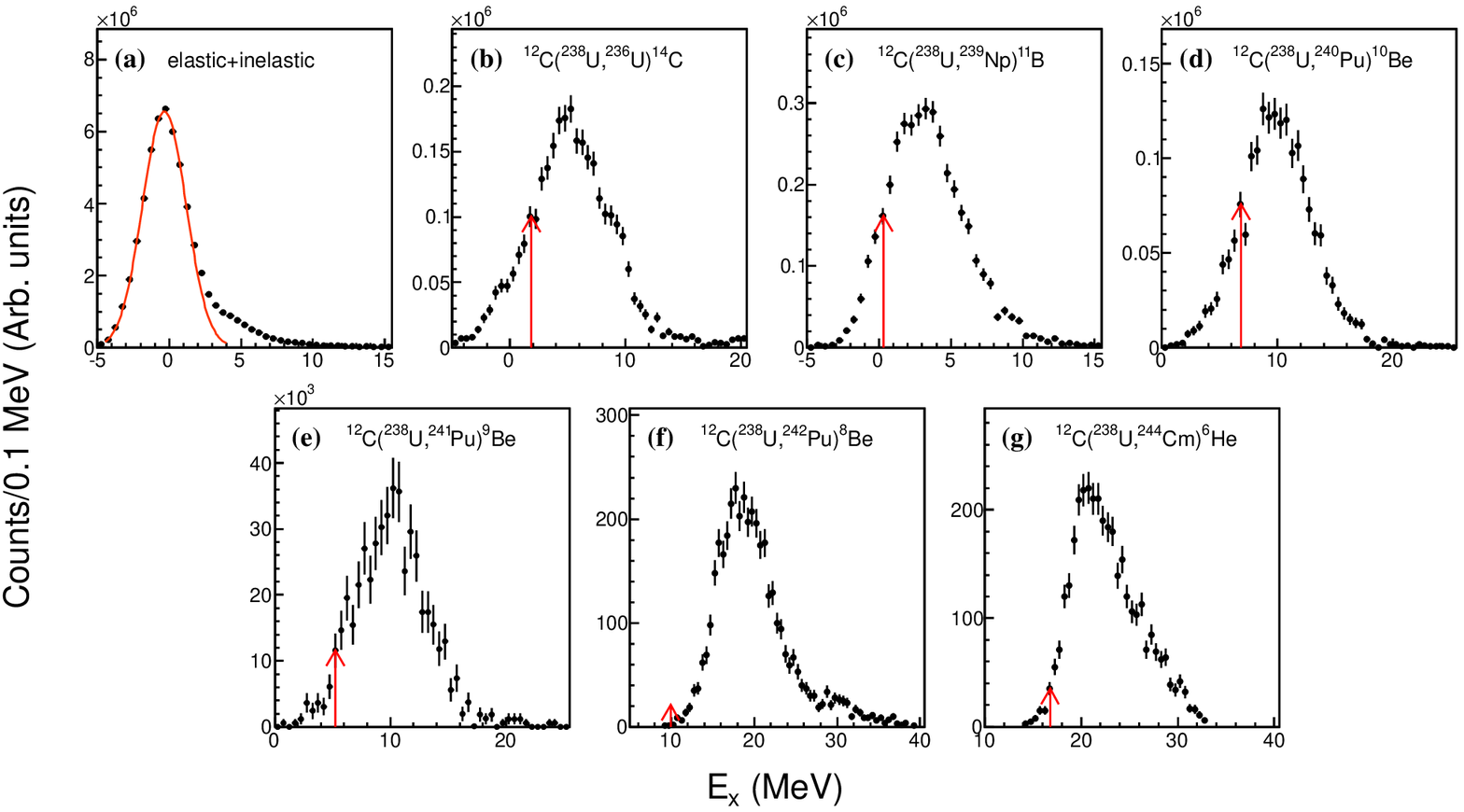}
\caption{\label{fig:Ex}(Color online) Total excitation-energy distributions measured for $^{238}$U+$^{12}$C elastic+inelastic scattering and multi-nucleon transfer reactions. The elastic peak was fitted to a Gaussian function in order to separate the inelastic component. The most probable excitation energies predicted by Eqs.~\ref{eq:Ex} and \ref{eq:Qopt} are indicated by arrows. The vertical error bars are statistical.}
\end{figure*}

\begin{figure}
\includegraphics[width=\linewidth]{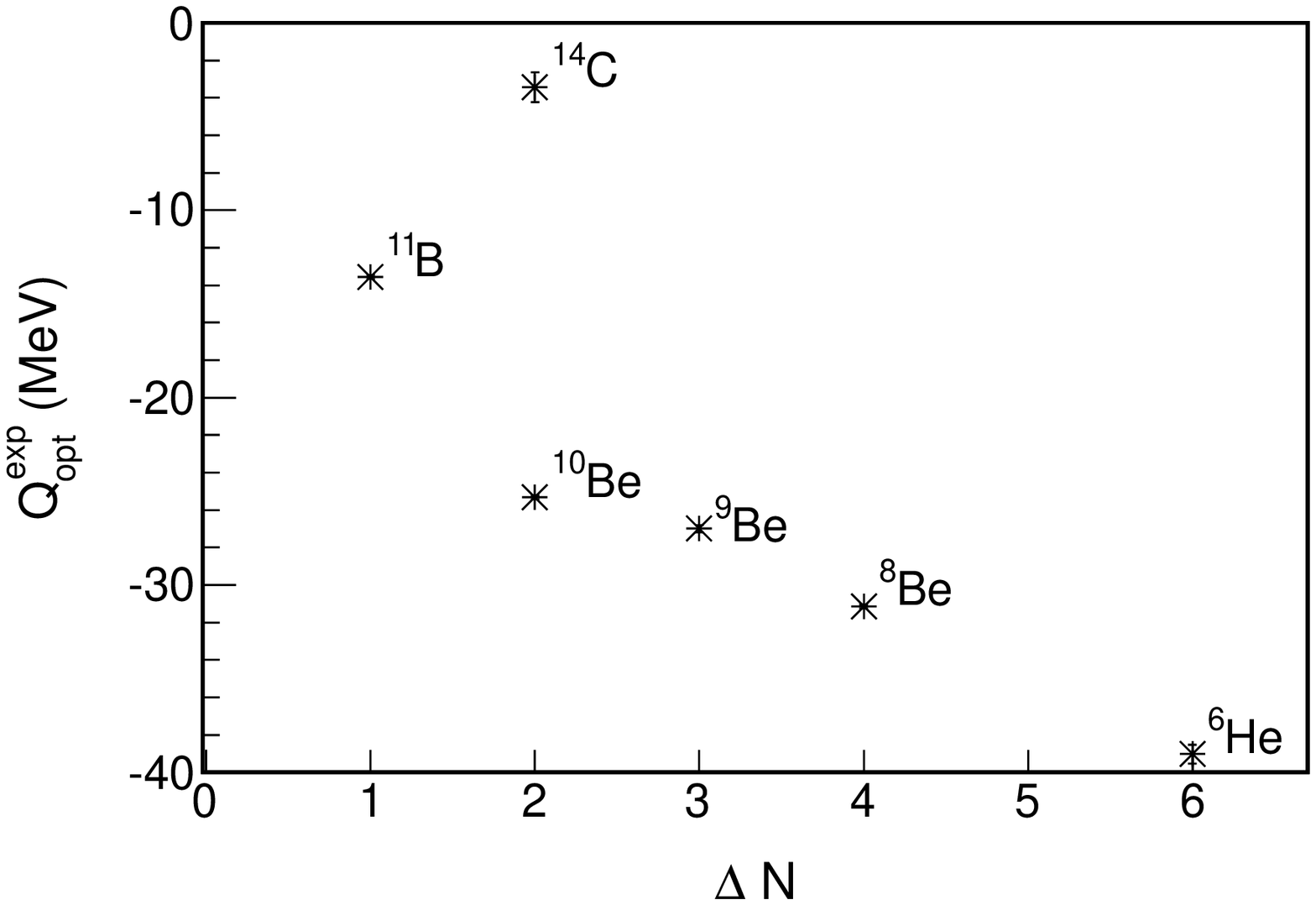}
\caption{\label{fig:Qopt}Optimal Q-values determined from the maxima of the total excitation energy distributions. Error bars are smaller than the point size.}
\end{figure}

The most probable total excitation-energy, $E_x$, in the exit channel can be obtained as the difference between the ground-state to ground-state and an effective Q-value, $Q_{opt}$, 

\begin{equation}
\begin{array}{c}
E_x=Q_{gg}-Q_{opt},\\
Q_{opt} = E_{i,c.m.}-E_{f,c.m.}.
\end{array}
\label{eq:Ex}
\end{equation}

In this expression, $E_{i,c.m.}$ and $E_{f,c.m.}$ are the initial and final total kinetic energies for a given transfer channel, in the center of mass.\\
If the transfer of nucleons is assumed to take place at the distance of closest approach, $D$, and collisions near the grazing angle are dominated by Coulomb forces, the orbit matching condition, $D_i=D_f$, leads to \cite{Alh79}

\begin{equation}
Q_{opt}= \frac{Z_3Z_4-Z_1Z_2}{Z_1Z_2}E_{i,c.m.},
\label{eq:Qopt}
\end{equation}

where $Z_{1,2,3,4}$, are the atomic numbers of the projectile, target, actinide and target-like nucleus, respectively. This expression is used in our work as reference, although it must be taken into account that no energy loss, i.e. $E_{i,c.m.}=E_{f,c.m.}$, is predicted when a mass transfer is not accompanied by a charge transfer.\\
In the present work, the total excitation energy available in the exit channel was derived from Eq.~\ref{eq:Ex}, using the energy and the angle of the target-like nucleus measured by the SPIDER telescope.\\
Figure \ref{fig:Ex} shows the reconstructed total excitation-energy distributions for $^{238}$U+$^{12}$C inelastic and multi-nucleon transfer reactions. The elastic peak was fitted to a Gaussian function in order to separate the inelastic component, which rapidly decreases with the excitation energy.\\
The total excitation-energy distributions populated in transfer reactions show a typical peaked shape, with a full width at half maximum of approximately 8 MeV. The maxima of the experimental distributions are systematically above the values given by Eqs.~\ref{eq:Qopt} and \ref{eq:Ex}, reflecting that optimal Q-values are smaller than the calculated ones. These results are shown in Table \ref{tab:xs_Ex}, which compiles the most probable excitation energies found in this work and the associated optimal Q-values. The first were obtained from Gaussian fits around the maxima of the distributions, which also provided the associated errors.\\
Figure \ref{fig:Qopt} shows that, contrary to the predictions of Eq.~\ref{eq:Qopt}, the amount of energy dissipated increases with the number of transferred nucleons, even if no charge exchange takes place. This result, which is in good agreement with Ref.~\cite{Kar82}, can be interpreted as an increase of the contact time when more nucleons are transferred. However, the transfer of two neutrons from the beam to the target, leading to $^{14}$C and $^{236}$U, considerably deviates from the reported trend, showing the influence of the Coulomb interaction in the kinetic energies of the emerging transfer partners, as it is expressed by Eq.~\ref{eq:Qopt}.\\
Finally, the incident neutron energies that, in a neutron capture reaction, would produce a compound nucleus with an excitation energy equal to the reported experimental values are also given in Table \ref{tab:xs_Ex}. Both quantities are related to each other through Eq.~\ref{eq:En}. The previous comparison assumes that the total excitation energy available from inelastic or transfer reactions is exclusively carried by the beam-like partner. Under this assumption, which will be discussed in Sec.~\ref{sec:gamma}, our data nicely show that transfer reactions cover different excitation-energy regimes. A wide region between equivalent thermal- and fast-neutron induced fission was sampled. The latter being of interest for fast-neutron reactor applications \cite{GEN}, where experimental data are strongly demanded \cite{Sal}.\\

\section{Excitation of light transfer partners}
\label{sec:gamma}

\begin{figure}	
{\includegraphics[width=\linewidth]{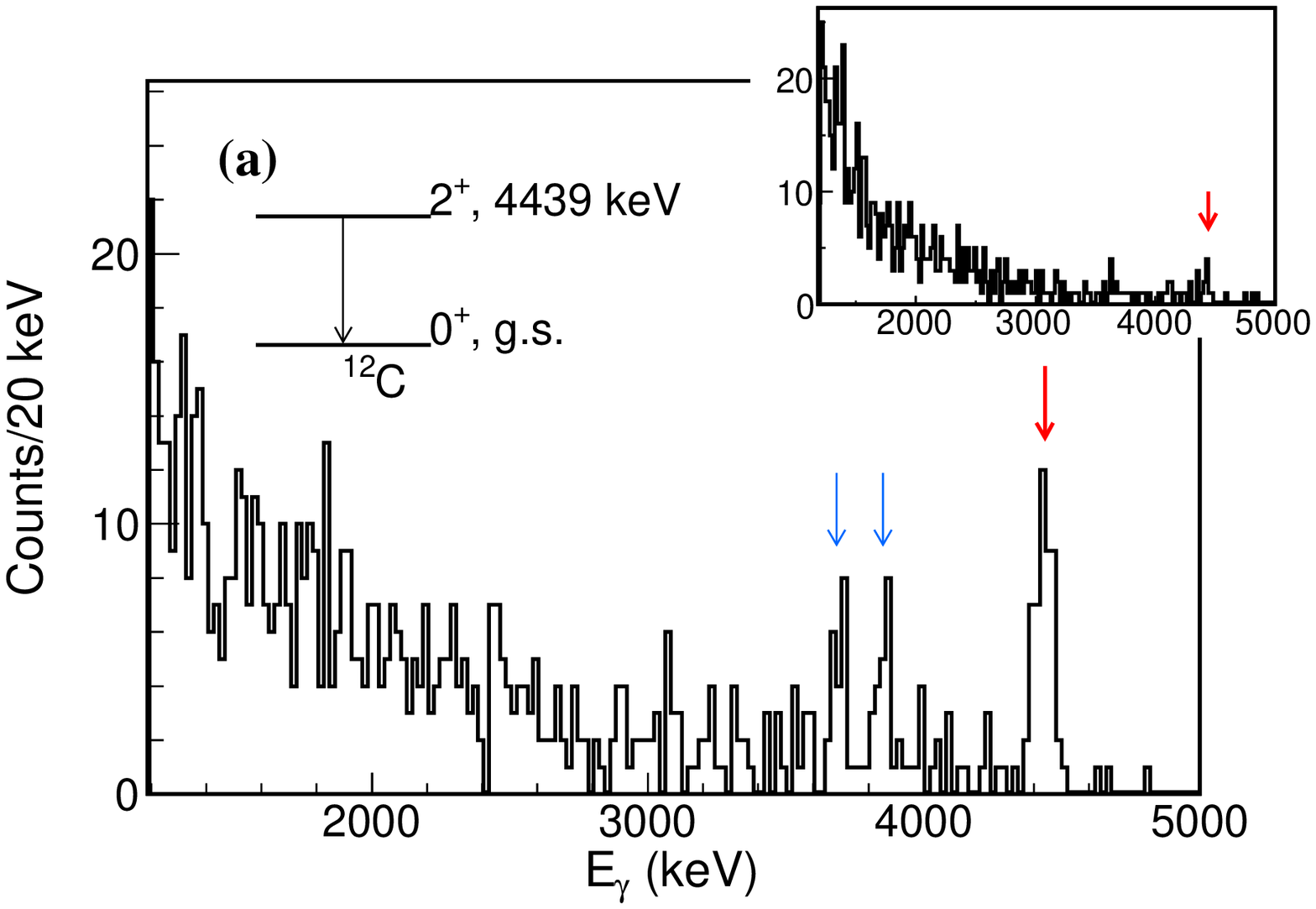}}
{\includegraphics[width=\linewidth]{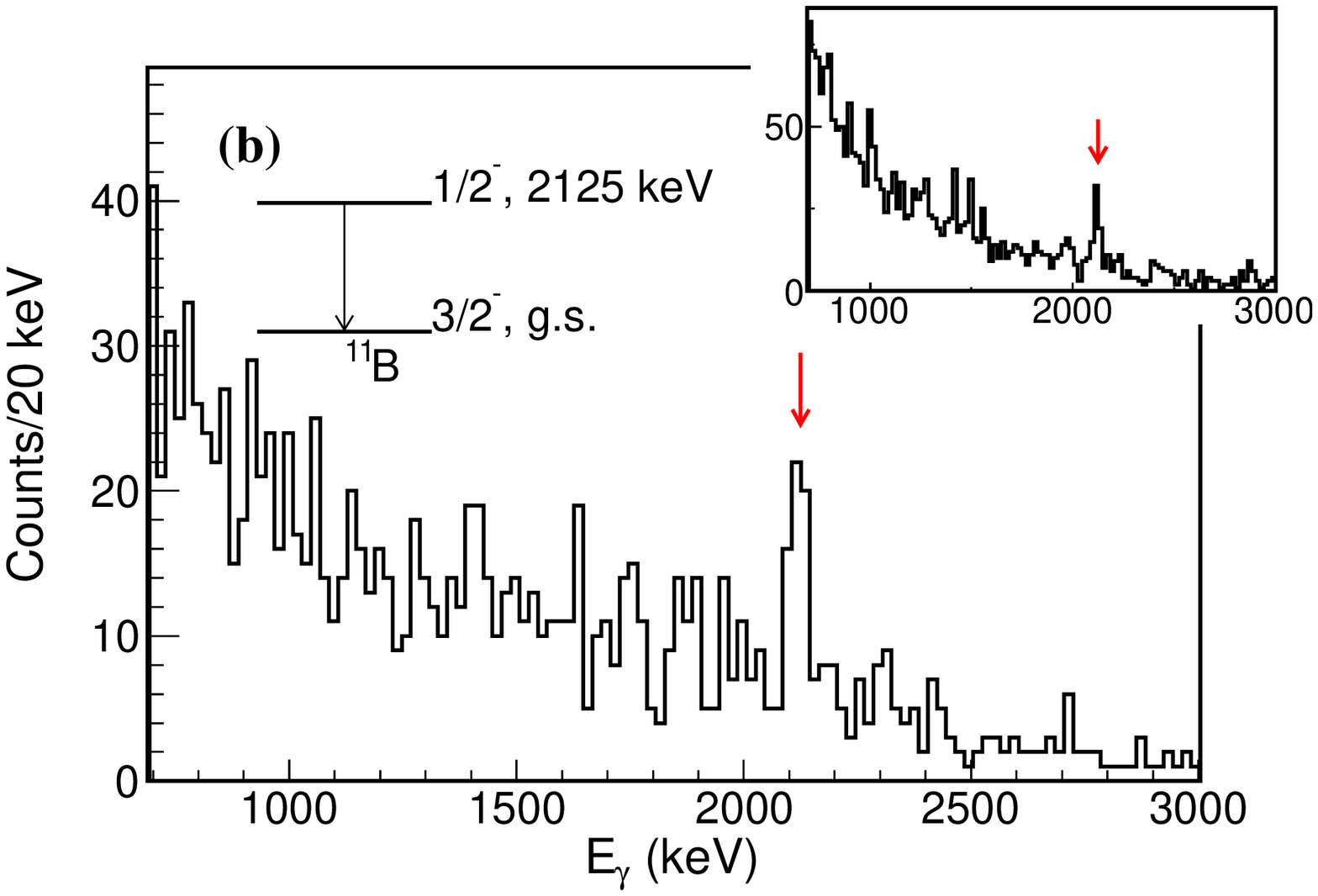}}
{\includegraphics[width=\linewidth]{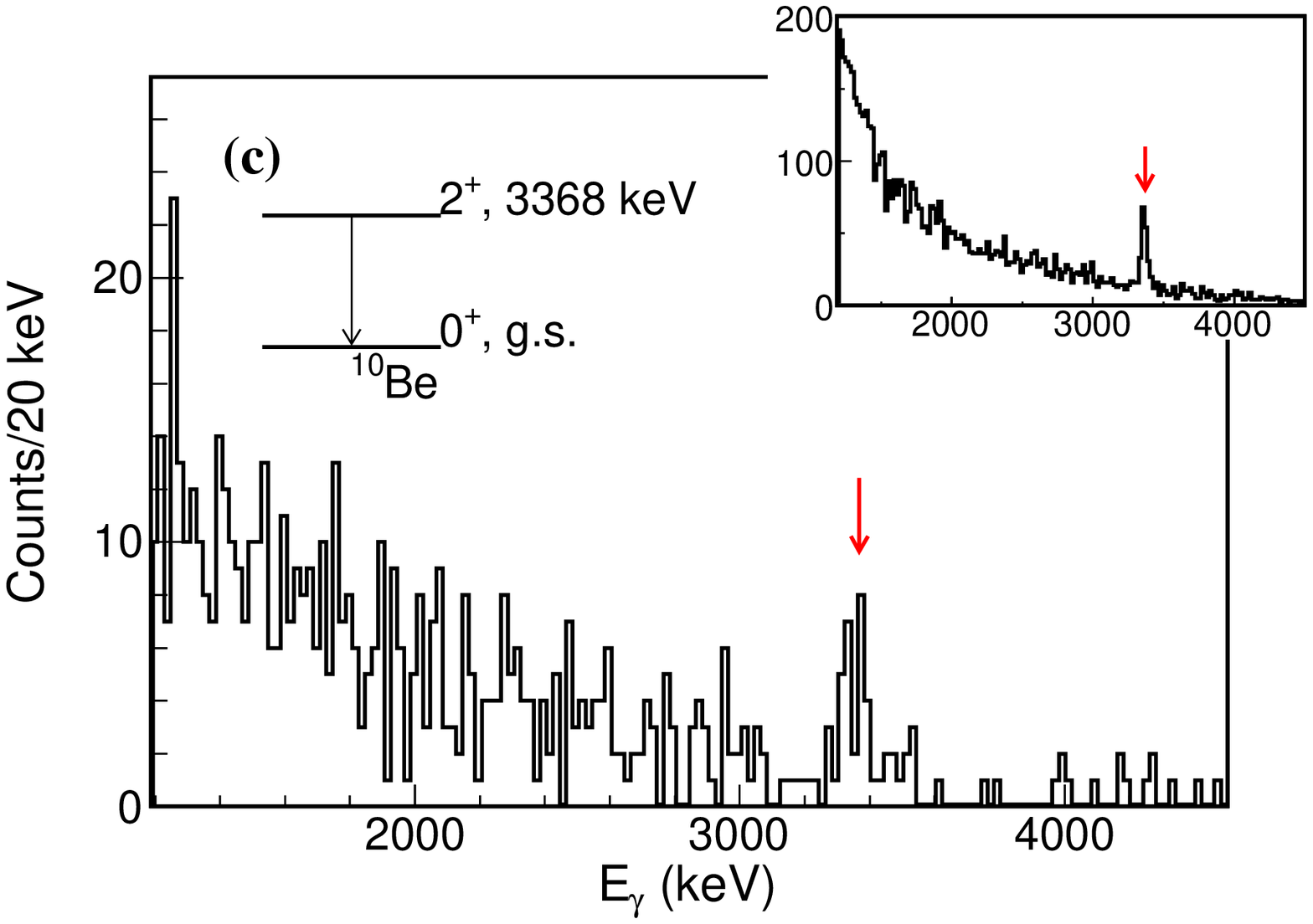}}
\caption{\label{fig:gamma}(Color online) Gamma-ray spectra measured for $^{12}$C (a), $^{11}$B (b) and $^{10}$Be (c) target-like nuclei. The diagrams therein represent the deexcitation from the first excited state. The associated $\gamma$ peaks are marked by thick (red) arrows. Thin (blue) arrows in the spectrum (a) show emissions from excited states of $^{13}$C. Main and inset plots correspond to trigger conditions that select anticoincidences and coincidences with the detection of a fission fragment in VAMOS, respectively.}
\end{figure}

\begin{table}
\caption{\label{tab:Pg}Probabilities of $\gamma$-ray emissions from the first excited states of target-like nuclei, $P_{\gamma}^{exp}$, determined in this work. The $\gamma$-ray energies, $E_{\gamma}$, and the associated detection efficiencies, $\epsilon_{\gamma}$, are also shown.}
\begin{ruledtabular}
\begin{tabular}{c c c c}
Target-like nucleus & E$_{\gamma}$ (keV) & $\epsilon_{\gamma}$ & P$_{\gamma}^{exp}$ \\
$^{12}$C & 4439 & 0.0059 $\pm$ 0.0004 & 0.14 $\pm$ 0.03 \\
$^{11}$B & 2125 & 0.0103 $\pm$ 0.0006 & 0.12 $\pm$ 0.02 \\
$^{10}$Be & 3368& 0.0073 $\pm$ 0.0004 & 0.14 $\pm$ 0.04 
\end{tabular}
\end{ruledtabular}
\end{table}

Possible excitations of the light inelastic and transfer partners were investigated by means of $\gamma$-ray measurements. Figure \ref{fig:gamma} shows the $\gamma$-ray spectra obtained for $^{12}$C, $^{11}$B and $^{10}$Be, in anticoincidence and coincidence (inset) with the detection of a fission fragment. A scaling factor of 600 should be applied in the first case, as only 1 of 600 anticoincidence events was accepted by the data-acquisition system during the experiment.\\
Clear $\gamma$-ray peaks were observed for $^{12}$C, $^{11}$B and $^{10}$Be, originating from the deexcitation of their first excited states. Regarding $^{11}$B and $^{10}$Be, similar results were found for events in anticoincidence and coincidence with the detection of a fission fragment. However, the characteristic $\gamma$-rays from the first excited state of $^{12}$C were hardly visible in fission events. This fact can be explained by the excitation-energy spectra populated in inelastic scattering. As it is shown in Fig.~\ref{fig:Ex}, the 10 MeV of excitation energy required to overcome the fission barrier of $^{238}$U and populate at the same time the first excited state of $^{12}$C are rarely reached.\\
No $\gamma$-ray emission was observed for $^{9}$Be nuclei, for which the first excited state is placed above the neutron-separation threshold. This result is in good agreement with the high probability of deexcitation through neutron emission reported for this state \cite{Til04}.\\
A quantitative analysis of the $\gamma$-energy spectra provided the probabilities of $\gamma$-ray emission from the first excited states of our target-like nuclei. They were calculated for each reaction channel as follows,

\begin{equation}
P_{\gamma} = \frac{ N_{f+\gamma}+D\cdot N_{\bar{f}+\gamma} }{ \epsilon_{\gamma}(N_{f}+D\cdot N_{\bar{f}}) }.
\end{equation}

The numbers of inelastic/transfer events in coincidence and  anticoincidence with the detection of a fission fragment in VAMOS are given by $N_{f}$ and $N_{\bar{f}}$, respectively, while $N_{f+\gamma}$ and $N_{\bar{f}+\gamma}$ correspond to the numbers of $\gamma$-rays observed in those events. The factor $D$ is the reduction applied in our acquisition system to $N_{\bar{f}}$ and $N_{\bar{f}+\gamma}$, $D=600$.\\
A minimum excitation energy was required for the computation of $N_{\bar{f}}$, which corresponds to the energy of the first excited state of the target like nucleus. This threshold was increased by the height of the actinide fission barrier for the computation of $N_{f}$, as this is the minimum excitation energy required for a simultaneous fission event. Finally, the parameter $\epsilon_{\gamma}$ represents the efficiency of our $\gamma$-ray measurements. Its determination was explained in Sec.~\ref{sec:analysis}.\\
The obtained probabilities are given in Table \ref{tab:Pg}. Similar results were found for $^{12}$C, $^{11}$B and $^{10}$Be, for which $P_{\gamma}=0.12-0.14$. A relative error between 15 and 20 \% must be considered for this result.\\
Those events where the target-like partner is excited require an additional total excitation energy in the exit channel to overcome the actinide fission barrier. A similar effect arises at the onset of the second-chance fission. As a consequence, the fission probability represented as a function of the total excitation energy is expected to differ from the results obtained in absence of target-like partner excitation and its average value will be slightly smaller.

\section{Fission probabilities}
\label{sec:Pf}

\subsection{Technical considerations}

For each transfer channel, the fission probabilities are defined as the proportion of events where the produced compound nuclei decay by fission. In our work, they can be derived as a function of the total excitation energy, $E_x$, using the following expression,

\begin{equation}
P_f(E_x) = \frac{N_{f}(E_x)}{a\cdot(N_{f}(E_x)+D \cdot N_{\bar{f}}(E_x))},
\label{eq:Pf}
\end{equation}

in which $N_{f}(E_x)$ and $N_{\bar{f}}(E_x)$ represent the number of events in coincidence and anticoincidence with the detection of a fission fragment in the VAMOS spectrometer. The total number of events for a given excitation energy is given by $N_{f}(E_x)+D \cdot N_{\bar{f}}(E_x)$, where $D$ is the reduction factor applied in our acquisition system to $N_{\bar{f}}$, $D=600$.\\
The term $a$ accounts for the acceptance of the VAMOS spectrometer. It was determined as the product of the acceptances in the azimutal an polar angles of the fission fragments, in the center of mass reference frame, denoted by $a_{\phi_{c.m.}}$ and $a_{\theta_{c.m.}}$ in the equation below.

\begin{equation}
a = a_{\phi_{c.m.}}\cdot a_{\theta_{c.m.}} 
\label{eq:acceptance}
\end{equation}

\begin{figure}
\includegraphics[width=\linewidth]{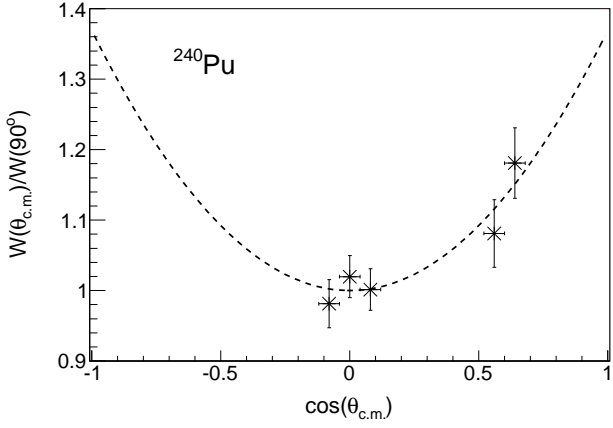}
\caption{\label{fig:anisotropy}Anisotropy of the fission fragments in transfer-induced fission of $^{240}$Pu. The curve is a fit of the experimental points. It follows Eq.~\ref{eq:ani}.}
\end{figure}

The azimutal and polar angle distributions were obtained by reconstructing the trajectories of the fission fragments through the VAMOS spectrometer \cite{Pul08}. A transformation into the center of mass reference frame was applied afterwards, using the kinematics of the transfer reaction.\\
Four configurations of VAMOS were used, in which the spectrometer was located at 20$^{\circ}$ and 14$^{\circ}$ with respect to the beam axis. The measurements at 20$^{\circ}$ were done with reference magnetic rigidities of 1.1 and 1.2 Tm, while 1.2 and 1.3 Tm were applied at 14$^{\circ}$. They allowed to scan polar angles of the fission fragments from 60$^{\circ}$ to 95$^{\circ}$, in the center of mass.\\
For each VAMOS configuration, the acceptance $a_{\phi_{c.m.}}$ was calculated as the ratio between the FWHM of the reconstructed $\phi_{c.m.}$ distribution and the total $2\pi$ range.\\
The calculation of $a_{\theta_{c.m.}}$ was based on the $\theta_{c.m.}$ distributions measured with the four VAMOS configurations. They were normalized by the acceptance in $\phi_{c.m.}$ and the beam intensity, which was accounted for by the number of elastic events detected in the SPIDER telescope. Then, few points around the maxima of these distributions were fitted to a function of the form:

\begin{equation}
W(\theta_{c.m.}) = W(90^{\circ})\cdot(1 + \alpha\cdot cos^2\theta_{c.m.}),
\label{eq:ani}
\end{equation}

providing a first-order description of the anisotropy of the fission fragments \cite{Van73}, $W(0^{\circ})/W(90^{\circ})$. An example is given in Fig.~\ref{fig:anisotropy}.\\
This procedure was applied to each transfer-induced fission channel. The resulting anisotropies are given in Table \ref{tab:PF}. The associated errors account for statistical fluctuations, as well as for movements of the beam of up to 2 mm influencing the rate of elastic events that we use for normalization. The uncertainty introduced by the simplification proposed in Eq.~\ref{eq:acceptance}, which neglects the interdependence between the magnetic rigidity, the polar and the azimuthal angles \cite{Caa13}, and the method given in Fig.~\ref{fig:anisotropy} was evaluated by means of a simulation and is included as well in the error bars.\\
For each configuration of VAMOS, the acceptance $a_{\theta_{c.m.}}$ was determined as the ratio between the integral of the $\theta_{c.m.}$ distribution and the integral of the anisotropy curve, in the range from 0$^{\circ}$ to 180$^{\circ}$. By combining these results with the previously calculated $a_{\phi_{c.m.}}$, total acceptances between 3 and 5 \% were obtained, depending on the derived anisotropy and the VAMOS configuration. The same error sources affect both the anisotropy and the acceptance calculations, although the latter are less sensitive to them, with a relative error $\delta_a/a = 0.2$. The final fission probabilities were obtained as the average of those measured with the four VAMOS configurations. \\
The influence of target-like nuclei scattered off the Al collimator on the excitation-energy distributions, which is discussed in Sec.~\ref{sec:analysis} and Fig.~\ref{fig:Ex_mask}, was observed in the $^{238}$U fission probabilities. Even though the particles scattered off the collimator were strongly suppressed by limiting the azimuthal and polar angles of target-like nuclei, a small fraction of elastic events wrongly reconstructed lead to a significant overestimation of $N_{\bar{f}} (E_{x})$, as the elastic cross-section dominates by an order of magnitude over the inelastic one. The quantity $N_{f}(E_x)$ remained comparably unaffected, because these events were always in coincidence with the detection of a fission fragment. As a result,  the fission probability was underestimated by approximately a factor of two. A correction factor was thus applied to the $^{238}$U fission-probability by scaling it to the distributions available from earlier works \cite{Cal80,Cra70}, using the points at 9 MeV of excitation energy as reference. However, the effect is still visible in the drop of the fission probability at excitation energies above 12 MeV. \\
A similar issue affected the $^{239}$Np fission probabilities, due to the wrong reconstruction of transfer events with excitation energies below the fission barrier. The scattering of these particles off the collimator lead to an overestimation of $N_{f} (E_{x})$, although with a more moderated effect because, in this case, the cross-section below the fission barrier is considerably smaller than for $^{238}$U. The restrictions applied in the azimuthal and polar angles of target-like nuclei provided $^{239}$Np fission probabilities in good agreement with previous data \cite{Gav76}, without need for a correction factor.

\subsection{Results}

\begin{figure*}
\includegraphics[width=\linewidth]{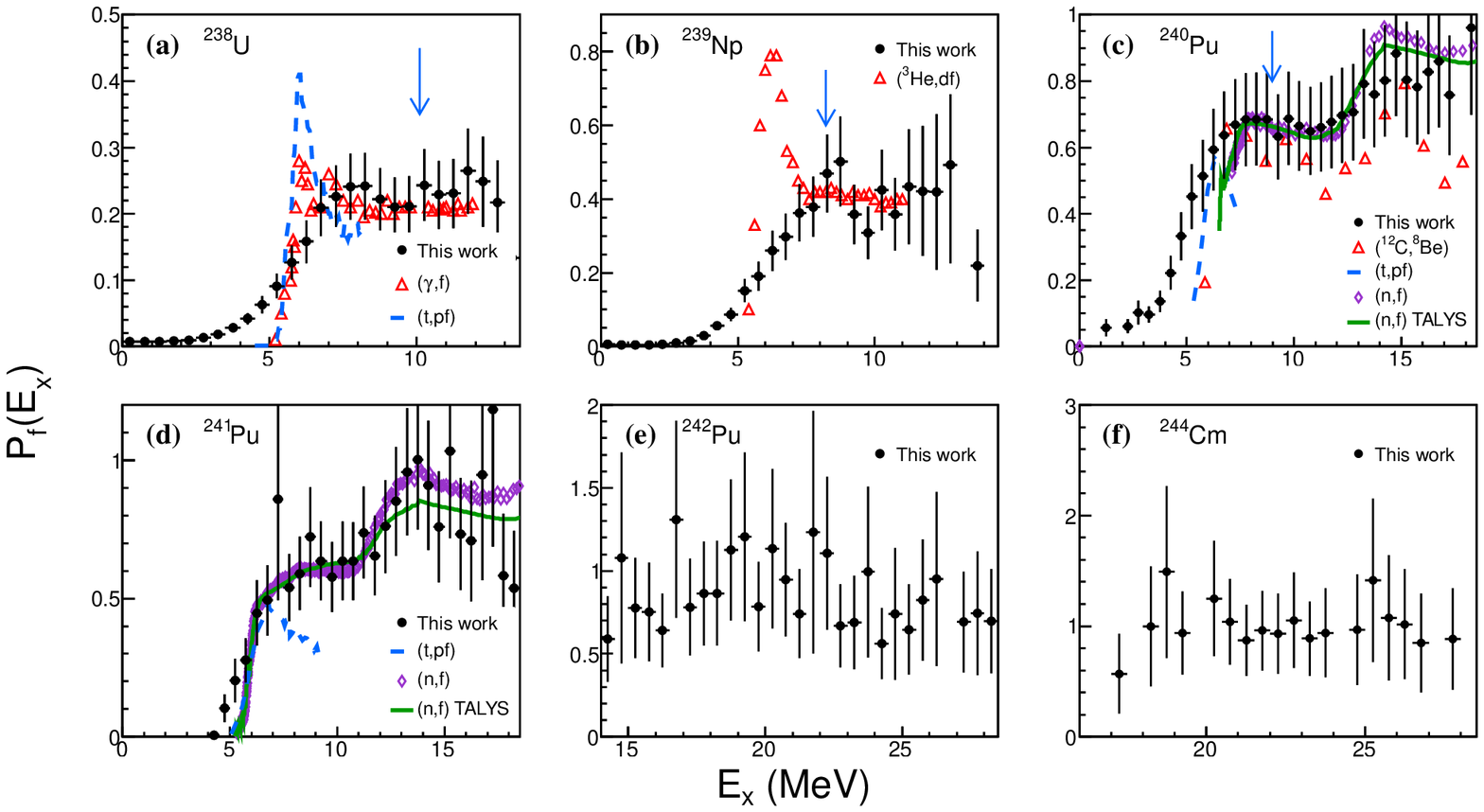}
\caption{\label{fig:PF}(Color online) Fission probabilities as a function of the total excitation energy. Results are presented for $^{238}$U+$^{12}$C inelastic scattering (a) and transfer-induced fission reactions (b-f). The fissioning nucleus is indicated in each figure. Earlier $\gamma$-, transfer- and neutron-induced fission data are included for comparison, as well as TALYS calculations of neutron-induced fission probabilities. The corresponding references are given in the text. Arrows point to excitation energies equal to the sum of the fission barrier and the energy of the first excited state of the light transfer partner.}
\end{figure*}

Figure \ref{fig:PF} shows the fission probabilities obtained in this work, as a function of the total excitation energy available in the exit channel. Six fissioning systems, between U and Cm, were investigated. The average values along the given excitation-energy energy ranges are compiled in Table \ref{tab:PF}.\\
The rising of $^{238}$U, $^{239}$Np and $^{240,241}$Pu fission probabilities at excitation energies of approximately 6 MeV shows the onset of first-chance fission, reflecting the heights of the fission barriers. The general agreement with earlier works and the recommended values of fission barriers \cite{Bjo80}, which are summarized in Table \ref{tab:PF}, brings confidence in the present technique, given the moderate excitation-energy resolution achievable in inverse-kinematics measurements.\\
Second-chance fission, for which a neutron is emitted before fission occurs, takes place at higher excitation energies. For reference, neutron separation energies are also reported in Table \ref{tab:PF}. This region could not be explored for $^{238}$U and $^{239}$Np because the minimum excitation energies required were barely reached, as shown in Fig.~\ref{fig:Ex}. However, the onset of second-chance fission is clearly visible for $^{240,241}$Pu, for which the fission probabilities suddenly increase when excitation energies of approximately 12 MeV are reached.\\
The fission of $^{242}$Pu and $^{244}$Cm was investigated at higher excitation energies. For the latter, a minor actinide of interest in accelerator-driven systems for the recycling of radioactive waste \cite{Art99}, the region above 15 MeV has been addressed for the first time \cite{Fur97}.\\
The comparison of our results with other works provides some hints concerning the influence of the reaction mechanism used to produce the fissioning system. One of these indications may be the disappearance of the peak near the onset of first chance fission for $^{238}$U and $^{239}$Np, which diverges from earlier ($\gamma$,f) \cite{Cal80}, (t,pf) \cite{Cra70} and ($^{3}$He,df) \cite{Gav76} measurements. The angular momentum transferred into the fissioning system in heavy-ion collisions may be inferred to explicate this difference.\\
Earlier transfer-induced fission measurements for $^{240,241}$Pu are limited to a single ($^{12}$C,$^{8}$Be)$^{240}$Pu experiment \cite{Che81} and ($t,pf$) reactions \cite{Cra70}, where only the region around the fission barrier was addressed. For both $^{240,241}$Pu isotopes, neutron-induced fission probabilities were calculated with the TALYS code \cite{TALYS}, as well the cross-sections of the formation of a compound nucleus in $^{239}$Pu+n and $^{240}$Pu+n reactions \cite{Rom}. They were corrected with a pre-equilibrium component, which becomes important at excitation energies above 10 MeV. Following Eq.~\ref{eq:surrogate}, the compound nucleus cross-sections were used to convert the experimental neutron-induced fission cross-sections provided in Ref.~\cite{Tov09} into fission probabilities. Figure \ref{fig:PF} (c) and (d) show the comparisons with our results.\\
Because the neutron separation energy of $^{240}$Pu is approximately 1 MeV above the fission barrier, excitation energies around the barrier can not be addressed in neutron capture reactions. The apparent threshold shown by $^{239}$Pu(n,f) data is actually related to the neutron-separation energy and reflects the importance of the so-called compound-nucleus elastic channel, where the compound nucleus decays to its ground state by emitting a neutron. A discrepancy of up to a factor of two is observed in this region with respect to our results. It may be explained by the higher angular momenta populated in our work, which forbids the compound-nucleus elastic branch.\\
The fission probabilities measured in ($^{12}$C,$^{8}$Be)$^{240}$Pu \cite{Che81} were obtained in a transfer reaction similar to the one that we use. However, these data present lower fission probabilities than measured in the present work. Surprisingly, they are also lower than the ones obtained in neutron-induced fission, whereas more angular momentum is supposed to be inferred in the fissioning system, increasing the fission probability. This fact may indicate an underestimation of the results given in Ref.~\cite{Che81}.\\
In the case of $^{241}$Pu, the fission barrier is above the neutron-separation energy, where the compound-nucleus elastic cross-section becomes less important. As a consequence, a better agreement with neutron-induced fission data is observed, both by our data and earlier ($t,pf$) measurements \cite{Cra70}.\\
A careful observation of Fig.~\ref{fig:PF} also provides some indications of the role played by excitations of the light transfer partner in the fission probabilities. Although tiny, our results for $^{238}$U, $^{239}$Np and $^{240}$Pu show structures at excitation energies corresponding to the sum of the height of the fission barrier and the energy of the first excited state of the light transfer partner, indicated by arrows in Fig.~\ref{fig:PF}. The are presumably produced by this effect.\\
The sensitivity of our results is limited by the background produced in the collimator behind the target, the excitation-energy resolution and the uncertainty in the determination of the acceptance of the VAMOS spectrometer, making it difficult to draw firmer conclusions on the interpretation of the measured fission probabilities and the differences with respect to the measurements in lighter transfer reactions. Further experiments of this type, as well as comparisons of the fission probabilities measured through different techniques, would be important in order to deepen into this discussion, which is crucial for the application of the surrogate-reaction technique. 

\begin{table*}
\caption{\label{tab:PF} Average fission probabilities obtained in this work, $\left<P_f\right>^{exp}$. For reference, the neutron separation energies, $S_n$, and the fission barriers \cite{Bjo80} are included. Because the fission barriers are double humped, they are given as $B_{f,A}$, and $B_{f,B}$.}
\begin{ruledtabular}
\begin{tabular}{c c c c c c c}
Actinide   & $S_n$ (MeV) & $B_{f,A}$ (MeV) & $B_{f,B}$ (MeV) & $W(0^{\circ})/W(90^{\circ})$ &$E_{x}$ range (MeV) & $\left<P_{f}\right>^{exp}$ \\
$^{238}$U  & 6.15 & 5.7 & 5.7 & 1.48$\pm$0.39 & 6--12 & 0.22$\pm$0.06\\
$^{239}$Np & 6.22 & 6.1 & 5.6 & 1.66$\pm$0.46 & 6--12 & 0.38$\pm$0.11\\
$^{240}$Pu & 6.53 & 5.6 & 5.1 & 1.37$\pm$0.36 & 5--18 & 0.72$\pm$0.23\\
$^{241}$Pu & 5.24 & 6.1 & 5.5 & 2.19$\pm$0.59 & 6--18 & 0.80$\pm$0.24\\
$^{242}$Pu & 6.31 & 5.6 & 5.1 & 1.00$\pm$0.28 & 14--28 & 0.86$\pm$0.33\\
$^{244}$Cm & 6.80 & 5.8 & 4.3 & 1.43$\pm$0.43 & 16--32 & 1.0$\pm$0.4 
\label{tab:PF}
\end{tabular}
\end{ruledtabular}
\end{table*}

\section{Summary and conclusions}
\label{sec:summary}

The present work approaches transfer-induced fission measurements from an innovative perspective, which implies the use of heavier transfer reactions and inverse kinematics. Inelastic and multi-nucleon transfer reactions between a $^{238}$U beam and a $^{12}$C target, at energies of approximately 10 \% above the Coulomb barrier, have been considered for the investigation of fission properties of exotic actinides, hardly accessible with standard neutron-irradiation measurements. As a result, up to nine different transfer channels were identified, leading to a variety of neutron-rich, excited actinides, from U to Cm.\\
The characterization of $^{238}$U+$^{12}$C inelastic and transfer channels was obtained from the detection and the identification of the final target-like partner in a Si telescope. Nucleons were mainly transferred from the light target to the heavy beam, while the flux in the opposite direction was limited to one and two neutrons. Cross-sections between few and few tens of mb were measured for the populated channels. The inelastic and transfer differential cross-sections were found to be bell-shaped and peaked above the calculated grazing angle.\\
In addition, total excitation-energy distributions were determined. The use of inverse kinematics, which makes possible the isotopic identification of the heavy and light fission fragments \cite{Caa13}, limited our excitation-energy resolution to 2.7 MeV (FWHM).\\
Our results show that higher excitation energies are reached as the number of nucleons transferred increases, even if no charge exchange takes places. The investigated reactions sample different excitation-energy regimes up to 30 MeV. In equivalent neutron-induced fission experiments, they would cover the region between thermal and fast neutron-induced fission.\\
Furthermore, possible excitations of the target-like nuclei in the exit channel were explored for the first time. The decays from the first excited states of $^{12}$C, $^{11}$B and $^{10}$Be were observed with probabilities of 0.12--0.14, by means of in-flight $\gamma$-ray measurements, performed in the target region. Beyond its importance for the characterization of transfer reactions, this result shows that the excitation of the light transfer-partner has to be taken into account for an accurate description of the fission probabilities within the surrogate-reaction technique.\\
Fission probabilities were determined for $^{238}$U, $^{239}$Np and $^{240,241}$Pu, as well as for $^{242}$Pu and $^{244}$Cm, where excitation energies above 15 MeV were addressed. The distributions obtained for the first reflect the height of the fission barrier, the role of the angular momentum populated in the fissioning system and the effect of the reported excitation of target-like nuclei, although the limited sensitivity of the results did not allow a deeper investigation of these features. Further investigations of this subject are advisable and will be of strong interest for the framework of the surrogate-reaction technique.\\
This work shows the potential of transfer-induced fission experiments with heavier targets, such as $^{12}$C, to access more neutron-rich and short-lived actinides, heavier than the beam. Future inverse-kinematics experiments based on this technique will benefit from the radioactive beams provided by ISOL facilities \cite{Her11,Lev11}, expanding the body of fissioning systems available to the experimental research.

\begin{acknowledgments}
The authors acknowledge the support from the GANIL staff during the experiment.\\
This work was partially supported by: Spanish Ministry of Education, under a fellowship for postdoctoral mobility (2011) administered by FECYT; Regional Council of Galicia, under a postdoctoral fellowship (2013); R\'egion Basse Normandie with a Chair of Excellence position in GANIL; EURATOM programme under contract 44816; EURATOM/FP7-249671; and GEDEPEON.
\end{acknowledgments}

\bibliography{biblio_SPIDER}

\end{document}